\documentclass[12pt]{article} 
\usepackage{makeidx}
\usepackage{amsthm,amsmath,amssymb,amsfonts}
\usepackage{thmtools}
\usepackage{setspace,graphicx,xcolor}
\usepackage{Generic}  
\usepackage[sort,longnamesfirst]{natbib}
\usepackage{url} 
\usepackage{algorithm}
\usepackage[noend]{algorithmic}
\usepackage{enumerate,enumitem}

\newcommand{\pcite}[1]{\citeauthor{#1}'s \citeyearpar{#1}}
\numberwithin{equation}{section}
\theoremstyle{plain}
\newtheorem{theorem}{Theorem}[section]
\declaretheorem[style=definition]{example}

\newcommand{\G}{{\mathsf{G}}}

\newcommand{\bb}{{\boldsymbol \beta}}

\numberwithin{equation}{section}

\allowdisplaybreaks

\begin{document}
%

\title{The data augmentation algorithm}

\author{Vivekananda Roy, Kshitij Khare and James P. Hobert}

 \date{}

 \maketitle

 \begin{abstract}
   The data augmentation (DA) algorithms are popular Markov chain
   Monte Carlo (MCMC) algorithms often used for sampling from
   intractable probability distributions. This review article
   comprehensively surveys DA MCMC algorithms, highlighting their
   theoretical foundations, methodological implementations, and
   diverse applications in frequentist and Bayesian statistics. The
   article discusses tools for studying the convergence properties of
   DA algorithms. Furthermore, it contains various strategies for
   accelerating the speed of convergence of the DA algorithms,
   different extensions of DA algorithms and outlines promising
   directions for future research. This paper aims to serve as a
   resource for researchers and practitioners seeking to
   leverage data augmentation techniques in MCMC algorithms by
   providing key insights and synthesizing recent developments.
\end{abstract}
\section{Introduction}
\label{sec:intro}
The data augmentation (DA) algorithm \citep{tann:wong:1987, swen:wang:1987} is a widely used class of Markov
chain Monte Carlo (MCMC) algorithms. Suppose $f_X$ is the target probability density function
(pdf) on $\mathbb{R}^p$. Generally, the goal is to evaluate the mean of
some function $h: \mathbb{R}^p \rightarrow \mathbb{R}$ with respect
to $f_X$, that is, to find
$\mbox{E}_{f_X}h = \int_{\mathbb{R}^p} h(x) f_X(x) \, dx$. Since $f_X$
is usually available only up to a normalizing constant,
$\mbox{E}_{f_X}h$ cannot be computed analytically. Furthermore, these
days, the target densities arising in machine learning, physics,
statistics, and other areas are so complex that direct simulation from
$f_X$ is impossible, making the classical Monte Carlo approximation of
$\mbox{E}_{f_X}h$ based on independent and identically distributed (iid) samples from $f_X$ infeasible.  In such situations, it is often
possible, as described later in a section, to construct a DA Markov
chain $\{ X_n \}_{n\ge 1}$, which has $f_X$ as its stationary density.
If this DA chain $\{ X_n \}_{n\ge 1}$ is well behaved then $\mbox{E}_{f_X}h$ can be
consistently estimated by the sample average $\bar{h}_n := \sum_{i=1}^n h(X_i)/n$.

For constructing a DA algorithm with stationary density $f_X$, one
needs to find a joint density $f(x, y)$ on
$\mathbb{R}^p \times \mathbb{R}^q$ with `augmented variables' $y$
satisfying the following two properties:
\begin{enumerate}
\item[(i)] the
  $x-$marginal of the joint density $f(x, y)$ is the target density $f_X$, that is,
 $\int_{\mathbb{R}^q} f(x,y) \, dy = f_X(x)$,  and
\item[(ii)] sampling from the two corresponding conditional
pdfs, $f_{X|Y}(x|y)$ and $f_{Y|X}(y|x)$ is straightforward. 
\end{enumerate}
The first property makes sure that the DA algorithm presented
below has $f_X$ as its stationary density, while the second property allows
for straightforward simulation of this DA Markov chain. As mentioned
in \cite{hobe:2011}, the popularity of the DA algorithm is due in part
to the fact that, given an intractable pdf $f_X$, there are some
general techniques available for constructing a potentially useful
joint density $f(x,y)$. For example, the augmented variables $y$
generally correspond to the \textit{missing data} used in the EM
algorithm \citep{demp:lair:rubi:1977} for defining the
\textit{complete data density}, which in turn is used for finding the
maximum likelihood estimates of some parameters. In
Sections~\ref{sec:exam}--\ref{sec:twoblock} we will provide several examples of widely used
nonlinear and/or high-dimensional statistical models with complex
target densities where using appropriate augmented data, popular and
efficient DA algorithms are constructed.

Each iteration of the DA algorithm consists of two steps --- a draw
from $f_{Y|X}$ followed by a draw from $f_{X|Y}$. Indeed, if the
current state of the DA Markov chain is $X_n = x$, we simulate $X_{n+1}$ as follows.

\begin{algorithm}[H]
  \caption{The $n$th iteration for the DA algorithm}
  \label{alg:da}
  \begin{algorithmic}[1]

    \vspace{.1in}
    
    \STATE Given $x$, draw $Y \sim f_{Y|X}(\cdot|x)$, and call the observed value $y$.

    \vspace{.1in}
    
    \STATE Draw $X_{n+1} \sim f_{X|Y}(\cdot|y)$.
    \vspace{.1in}
    
\end{algorithmic}
\end{algorithm}
From the two steps in each iteration of the DA algorithm, it follows that the Markov transition
density (Mtd) of the DA Markov chain $\{ X_n
\}_{n\ge 1}$ is given by
\begin{equation}
  \label{eq:da_mtd}
  k(x'|x) = \int_{\mathbb{R}^q} f_{X|Y}(x'|y) f_{Y|X}(y|x) \, dy \;.
\end{equation}
Let $f_Y(y) = \int_{\mathbb{R}^p} f(x,y) \, dx$ be the marginal
density of the augmented variable $y$. Since
\begin{equation}
  \label{eq:dbal}
  k(x'|x) f_X(x) = f_X(x) \int_{\mathbb{R}^q} f_{X|Y}(x'|y) f_{Y|X}(y|x) \, dy =
  \int_{\mathbb{R}^q} \frac{f(x',y) f(x,y)}{f_Y(y)} \, dy = k(x|x') f_X(x'),
\end{equation}
for all $x,x'$, that is, the Mtd $k$ satisfies the \textit{detailed balance
  condition}, we have
\begin{equation}
\label{eq:stationary}
\int_{\mathbb{R}^p} k(x'|x) f_X(x) \, dx = f_X(x') \;.
\end{equation}
Thus, $f_X$ is stationary for the DA Markov chain
$\{ X_n \}_{n\ge 1}$. Moreover, if $\{ X_n \}_{n\ge 1}$ is
appropriately irreducibile, then $\bar{h}_n$ is a strongly consistent
estimator of $\mbox{E}_{f_X}h$, that is, $\bar{h}_n$ converges to
$\mbox{E}_{f_X}h$ almost surely as $n \rightarrow \infty$ \citep[Theorem
2]{asmu:glyn:2011}. Since the DA chain has the Mtd \eqref{eq:da_mtd}, irreducibility also implies Harris recurrence of the DA Markov chain. Furthermore, if $\{ X_n \}_{n\ge 1}$ is also
aperiodic then it is \textit{Harris ergodic} and the marginal density
of $X_n$ converges to the stationary density $f_X$, no matter what the
initial distribution of $X_1$ is \citep[][chapter 13]{meyn:twee:1993}. A
simple sufficient condition which guarantees both irreducibility and
aperiodicity of $\{ X_n \}_{n\ge 1}$ is that the Mtd $k$ is strictly
positive everywhere. Note that, if $f(x, y)$ is strictly positive on $\mathbb{R}^p \times \mathbb{R}^q$,
then $k$ is strictly positive everywhere and the DA Markov chain $\{ X_n \}_{n\ge 1}$ is Harris ergodic.


In order to keep the notations and discussions simple, here we
consider the target and augmented state space as Euclidean. Similarly,
the densities $f_X$ as well as $f(x, y)$ are considered with respect
to the Lebesgue measure. However, all methods and theoretical results
discussed in this chapter extend to discrete as well as other general
settings. Indeed, there are popular DA algorithms where the target
and/or the augmented state space is not continuous, for example
consider the widely used DA algorithm for the mixture models \citep{dieb:robe:1994}.



\section{Examples}
\label{sec:exam}
In this section, we provide examples of some widely used high
dimensional linear models, generalized linear models, and generalized
linear mixed models where the target distributions are intractable. Then we
describe some DA algorithms that are used to fit these models.

\subsection{Data augmentation for Bayesian variable selection}
\label{sec:bvs}
In modern datasets arising from diverse applications such as
genetics, medical science, and other scientific disciplines, the number
of covariates are often larger than the number of observed samples. The
ordinary least squares method for estimating the regression
coefficients in a linear regression is not applicable in such situations. On the other hand,
penalized regression methods such as the {\it least absolute shrinkage
  and selection operator} ({\it lasso}) \citep{tibs:1996} are
applicable as they can simultaneously induce shrinkage and sparsity
in the estimation of the regression coefficients.

Consider the standard linear model
\[
  Z|\mu, \beta, \sigma^2 \sim N_m(\mu  1_m + W \beta, \sigma^2I_m),
\]
where $Z \in \mathbb{R}^m$ is the vector of responses,
$\mu \in \mathbb{R}$ is the intercept, $ 1_m$ is the $m \times 1$
vector of 1's, $W$ is the $m \times p$ ({\it standardized}) covariate
matrix, $\beta = (\beta_1,\dots,\beta_p) \in \mathbb{R}^p$ is the
vector of regression coefficients, and $\sigma^2$ is the variance
parameter. The lasso is based on $L_1$ norm regularization, and it
estimates $\beta$ by solving
\begin{equation}
  \label{eq:lass}
  \underset{\beta}{\mbox{min}}\; (\tilde{z} - W\beta)^\top(\tilde{z} - W\beta) + \lambda \sum_{j=1}^p |\beta_j|,
\end{equation}
for some shrinkage parameter $\lambda \in \mathbb{R}$, where
$z$ is the vector of observed responses, and $\tilde{z} = z - {\bar z} 1_m$. The lasso estimate can be interpreted
as the posterior mode when, conditional on $\sigma^2$, the regression parameters $\beta_j$'s have
independent and identical Laplace priors \citep{tibs:1996}. Indeed, following
\cite{park:case:2008}, we consider the hierarchical model
\begin{eqnarray}
\label{eq:lassmodel}
 Z| \mu, \beta, \sigma^2 &\sim& N_m(\mu  1_m + W \beta, \sigma^2I_m),\nonumber\\
\pi(\mu) \propto 1; \beta | \sigma^2 &\sim& \prod_{j=1}^p \frac{\lambda}{2\sqrt{\sigma^2}}e^{-\lambda |\beta_j|/\sqrt{\sigma^2}}
\nonumber\\
\sigma^2 &\sim& \pi(\sigma^2),
\end{eqnarray}
where $\pi(\sigma^2)$ is the prior density of $\sigma^2$.  Since the
columns of $W$ are centered, standard calculations show that
\begin{equation}
  \label{eq:intmu}
  f(z | \beta, \sigma^2) \equiv \int_{\mathbb{R}} f(z
| \mu, \beta, \sigma^2) \pi(\mu) d\mu = \frac{1}{(2\pi)^{(n-1)/2}
  \sigma^{n-1}} \exp\Big[-\frac{(\tilde{z} - W \beta)^\top(\tilde{z} - W \beta)}{2\sigma^2}\Big].
\end{equation}
Thus, from \eqref{eq:lassmodel} and \eqref{eq:intmu}, the joint posterior density of $(\beta, \sigma^2)$ is
\begin{equation}
  \label{eq:lasstarget}
  \pi(\beta, \sigma^2| z) \propto \frac{1}{(\sigma^2)^{(n-1+p)/2}} \exp\Bigg[-\frac{(\tilde{z} - W \beta)^\top(\tilde{z} - W \beta)+ 2\lambda\sigma \sum_{j=1}^p |\beta_j|}{2\sigma^2}\Bigg] \pi(\sigma^2).
\end{equation}
From \eqref{eq:lass} and \eqref{eq:lasstarget} it follows that the mode of the
(conditional on $\sigma^2$) posterior density of $\beta$ is the lasso estimate.

The posterior density \eqref{eq:lasstarget} is intractable. Introducing augmented
variables $y=(y_1,\dots,y_p)$ with $y_i >0$ for all $i$, \cite{park:case:2008} consider the following joint density
of $(\beta, \sigma^2, y)$
\begin{align}
  \label{eq:lasscomp}
  \pi(\beta, \sigma^2, y| z) &\propto \frac{1}{(\sigma^2)^{(n-1+p)/2}} \exp\Bigg[-\frac{(\tilde{z} - W \beta)^\top(\tilde{z} - W \beta)}{2\sigma^2}\Bigg] \nonumber\\
  &\hspace{1in} \times \Bigg[\prod_{j=1}^p \frac{1}{\sqrt{y_j}}\exp\bigg\{- \frac{\beta_j^2}{2 \sigma^2y_j}\bigg\} \exp\bigg\{- \frac{\lambda^2y_j}{2}\bigg\}\Bigg]\pi(\sigma^2).
\end{align}
Replacing $t$ with $\beta_j/\sigma$, $s$ with $y_j$, and $a$ with $\lambda$ in the following representation of the Laplace density as a scale mixture of normals \citep{andr:mall:1974}
\begin{equation}
  \label{eq:mixiden}
  \frac{a}{2} \exp(-a |t|) = \int_0^{\infty} \frac{1}{\sqrt{2 \pi s}} \exp\Big(- \frac{t^2}{2s}\Big) \frac{a^2}{2} \exp\Big(- \frac{a^2}{2} s\Big) ds, \; a>0
\end{equation}
we see that
\[
  \int_{\mathbb{R}^p} \pi(\beta, \sigma^2, y| z) dy = \pi(\beta, \sigma^2| z),
  \]
  that is, the $(\beta, \sigma^2)-$ marginal density of the joint
  posterior density $\pi(\beta, \sigma^2, y| z)$ given in
  \eqref{eq:lasscomp} is the target posterior density
  \eqref{eq:lasstarget}.  Thus, from Section~\ref{sec:intro}, if
  sampling from the two conditional densities
  $\pi(\beta, \sigma^2| y, z)$ and $\pi(y| \beta, \sigma^2, z)$ is
  straightforward, then we can construct a valid DA algorithm for
  \eqref{eq:lasstarget}.

  From \eqref{eq:lasscomp}, we have
  \begin{equation}
    \label{eq:lasscondy}
        \pi(y| \beta, \sigma^2, z) \propto \prod_{j=1}^p \frac{1}{\sqrt{y_j}}\exp\bigg\{- \frac{\beta_j^2}{2 \sigma^2y_j}\bigg\} \exp\Big\{- \frac{\lambda^2y_j}{2}\Big\}.
  \end{equation}
    Recall that the Inverse-Gaussian $(\kappa, \psi)$ density is given by
    \[
      f(u) = \sqrt{\frac{\psi}{2\pi}} u^{-\frac{3}{2}} \exp\Big\{-\frac{\psi(u-\kappa)^2}{2\kappa^2u}\Big\}.
    \]
    Thus from \eqref{eq:lasscondy}, it follows that
    \begin{equation}
      \label{eq:lasscondydis}
     \frac{1}{y_j} \Big| \beta, \sigma^2, z \stackrel{ind}\sim \mbox{Inverse-Gaussian} \Big(\sqrt{\frac{\lambda^2 \sigma^2}{\beta^2_j}}, \lambda^2\Big)\;\mbox{for}\; j=1,2,\dots,p.
   \end{equation}

   In order to derive the conditional density
   $\pi(\beta, \sigma^2| y, z)$ we assume that apriori
   $\sigma^2 \sim \mbox{Inverse-Gamma} (\alpha, \xi)$ for some
   $\alpha \ge 0$ and $\xi \ge 0$. The improper prior
   $\pi(\sigma^2) = 1/\sigma^2$ used in \cite{park:case:2008} is
   obtained by replacing $\alpha=0, \xi=0$. Note that a draw from
   $\pi(\beta, \sigma^2| y, z)$ can be made by first drawing from
   $\pi(\sigma^2| y, z)$ followed by a draw from
   $\pi(\beta| \sigma^2, y, z)$ \citep{raja:spar:khar:zhan:2019}. Denoting the $p \times p$ diagonal matrix with diagonal elements
   $(y_1,\dots, y_p)$ by $D_y$, we have
\begin{align*}
     \pi(\beta| \sigma^2, y, z) &\propto \pi(\beta, \sigma^2, y| z) \propto \exp\Bigg[-\frac{(\tilde{z} - W \beta)^\top(\tilde{z} - W \beta) + \beta^\top D_y^{-1} \beta}{2\sigma^2}\Bigg]\\
     &\propto \exp\Bigg[-\frac{\beta^\top(W^{\top}W + D_y^{-1})\beta -2 \beta^{\top} W^{\top}\tilde{z}}{2\sigma^2}\Bigg].
   \end{align*}
   Thus,
   $ \beta | \sigma^2, y, z \sim N_p ((W^\top W + D_y^{-1})^{-1}
   W^{\top} \tilde{z}, \sigma^2(W^{\top}W + D_y^{-1})^{-1})$. Also,
   from \eqref{eq:lasscomp}, we have
   \begin{align*}
     \pi(\sigma^2| y, z) &\propto \pi(\sigma^2, y| z) = \int_{\mathbb{R}^p} \pi(\beta, \sigma^2, y| z) d\beta\\
                         &\propto \int_{\mathbb{R}^p} \frac{1}{(\sigma^2)^{(n-1+p)/2}} \exp\Bigg[-\frac{(\tilde{z} - W \beta)^\top(\tilde{z} - W \beta) + \beta^\top D_y^{-1} \beta}{2\sigma^2}\Bigg] \pi(\sigma^2) d\beta\\
                              &\propto \frac{\pi(\sigma^2)}{(\sigma^2)^{(n-1)/2}} \exp \Big(-\frac{\tilde{z}^\top\tilde{z}}{2\sigma^2}\Big) \int_{\mathbb{R}^p} (\sigma^2)^{-p/2} \exp\Bigg[-\frac{\beta^\top(W^{\top}W + D_y^{-1})\beta -2 \beta^{\top} W^{\top}\tilde{z}}{2\sigma^2}\Bigg]  d\beta\\
     &\propto (\sigma^2)^{-(n-1 +2\alpha)/2 -1} \exp\Bigg[-\frac{\tilde{z}^\top(I - W(W^{\top}W + D_y^{-1})^{-1}W^{\top})\tilde{z}+2\xi}{2\sigma^2}\Bigg].
   \end{align*}
Thus, $\sigma^2|y, z \sim \mbox{Inverse-Gamma} ( (n-1+2\alpha)/2, (\tilde{z}^\top(I - W(W^{\top}W + D_y^{-1})^{-1}W^{\top})\tilde{z}+2\xi)/2).$ Hence, a single iteration of the DA algorithm uses the following
two steps to move from $(\beta, \sigma^2)$ to $(\beta', {\sigma^2}')$.
\begin{algorithm}[H]
    \caption{One iteration of the DA algorithm for Bayesian lasso}
    \begin{algorithmic}[1]
          \STATE Given $(\beta, \sigma^2)$, draw $y_1,\dots, y_p$ independently with
          \[
            \frac{1}{y_j} \Big| \beta, \sigma^2, z\sim \mbox{Inverse-Gaussian} \bigg(\sqrt{\frac{\lambda^2 \sigma^2}{\beta^2_j}}, \lambda^2\bigg)\;\mbox{for}\; j=1,2,\dots,p.
          \]

                        \STATE Draw $(\beta^\prime, {\sigma^2}^\prime)$ by first drawing
                        \[
                          {\sigma^2}^\prime|y, z \sim \mbox{Inverse-Gamma} \Big( \frac{n-1}{2} + \alpha, \frac{\tilde{z}^\top(I - W(W^{\top}W + D_y^{-1})^{-1}W^{\top})\tilde{z}}{2}+\xi \Big),
                          \]
 and then drawing

              \[
                {\beta}^\prime | {\sigma^2}^\prime, y, z \sim N_p ((W^\top W + D_y^{-1})^{-1}
   W^{\top} \tilde{z}, {\sigma^2}^\prime(W^{\top}W + D_y^{-1})^{-1}).
                \]

	\end{algorithmic}
      \end{algorithm}
      
      Although the lasso estimator has been extensively used in
      applications as diverse as agriculture, genetics, and finance,
      it may perform unsatisfactorily if the predictors are highly
      correlated. For example, if there is a group structure among the
      variables, lasso tends to select only one variable from each
      group. \cite{zou:hast:2005} proposed the {\it Elastic Net} (EN)
      to achieve better performance in such situations. The EN
      estimator is obtained by solving
\begin{equation}
  \label{eq:en}
  \underset{\beta}{\mbox{min}}\; (\tilde{z} - W\beta)^\top(\tilde{z} - W\beta) + \lambda_1 \sum_{j=1}^p |\beta_j| + \lambda_2 \sum_{j=1}^p |\beta_j|^2,
\end{equation}
where $\lambda_1$ and $\lambda_2$ are tuning parameters. From
\eqref{eq:en} we see that the elastic net uses both an $L_1$ penalty
as in lasso and an $L_2$ penalty as in the ridge regression. Following
\cite{kyun:gill:ghos:case:2010} and \cite{roy:chak:2017}, we consider
the hierarchical Bayesian EN model
\begin{eqnarray}
\label{eq:enmodel}
  Z| \mu, \beta, \sigma^2 &\sim& N_m(\mu  1_m + W \beta, \sigma^2I_m),\nonumber\\
  \pi(\mu) \propto 1;   \pi(\beta | \sigma^2) &\propto& \prod_{j=1}^p \frac{1}{\sigma}\exp\Big\{ -\frac{\lambda_1
                                                        |\beta_j|}{\sigma} - \frac{\lambda_2 \beta^2_j}{2 \sigma^2}\Big\}, 
                                                        \nonumber\\
  \sigma^2 &\sim& \pi(\sigma^2),
\end{eqnarray}
where $\pi(\sigma^2)$ is the prior density of $\sigma^2$. From
\eqref{eq:enmodel} and \eqref{eq:intmu} it follows that the joint posterior density
of $(\beta, \sigma^2)$ is
\begin{equation}
  \label{eq:entarget}
  \pi_{\text{EN}}(\beta, \sigma^2| z) \propto \frac{\pi(\sigma^2)}{(\sigma^2)^{(n-1+p)/2}} \exp\Bigg[-\frac{(\tilde{z} - W \beta)^\top(\tilde{z} - W \beta)+ 2\lambda_1\sigma \sum_{j=1}^p |\beta_j| + \lambda_2 \sum_{j=1}^p \beta_j^2}{2\sigma^2}\Bigg].
\end{equation}
From \eqref{eq:en} and \eqref{eq:entarget} we see that the mode of the
(conditional on $\sigma^2$) posterior density of $\beta$ is the EN
estimate. Since the density \eqref{eq:entarget} is intractable, using
augmented variables $y=(y_1,\dots,y_p)$, following
\cite{kyun:gill:ghos:case:2010} and \cite{roy:chak:2017}, we consider
the joint density of $(\beta, \sigma^2, y)$ given by
\begin{align}
  \label{eq:encomp}
  \pi_{\text{EN}}(\beta, \sigma^2, y| z) &\propto \frac{\pi(\sigma^2)}{(\sigma^2)^{(n-1+p)/2}} \exp\Bigg[-\frac{(\tilde{z} - W \beta)^\top(\tilde{z} - W \beta)}{2\sigma^2}\Bigg] \nonumber\\
  &\hspace{.5in} \times \Bigg[\prod_{j=1}^p \frac{1}{\sqrt{y_j}}\exp\bigg\{- \frac{(\lambda_2+1/y_j) \beta_j^2}{2 \sigma^2}\bigg\} \exp\bigg\{- \frac{\lambda_1^2y_j}{2}\bigg\}\Bigg].
\end{align}
From \eqref{eq:mixiden} we see that
\[
  \int_{\mathbb{R}^p} \pi_{\text{EN}}(\beta, \sigma^2, y| z) dy = \pi_{\text{EN}}(\beta, \sigma^2| z).
\]
From \eqref{eq:encomp} using similar calculations as before, we see
that the conditional distributions of $1/y_j$'s are the same as
\eqref{eq:lasscondydis} with $\lambda=\lambda_1$. As before, we assume that apriori
   $\sigma^2 \sim \mbox{Inverse-Gamma} (\alpha, \xi)$ for some
   $\alpha \ge 0$ and $\xi \ge 0$. Denoting the $p \times p$ diagonal matrix with diagonal elements
   $(1/(\lambda_2+1/y_1),\dots, 1/(\lambda_2+1/y_p))$ by $\tilde{D}_y$, we have
\begin{align*}
     \pi_{\text{EN}}(\beta| \sigma^2, y, z) \propto \pi_{\text{EN}}(\beta, \sigma^2, y| z) \propto \exp\Bigg[-\frac{(\tilde{z} - W \beta)^\top(\tilde{z} - W \beta) + \beta^\top \tilde{D}_y^{-1} \beta}{2\sigma^2}\Bigg].
   \end{align*}
   Thus,
   $ \beta | \sigma^2, y, z \sim N_p ((W^\top W + \tilde{D}_y^{-1})^{-1}
   W^{\top} \tilde{z}, \sigma^2(W^{\top}W + \tilde{D}_y^{-1})^{-1})$.
Next, doing similar calculations as for the Bayesian lasso, we see that $\sigma^2|y, z \sim \mbox{Inverse-Gamma} ( (n-1+2\alpha)/2, (\tilde{z}^\top(I - W(W^{\top}W + \tilde{D}_y^{-1})^{-1}W^{\top})\tilde{z}+2\xi)/2).$ Hence, a single iteration of the DA algorithm for the Bayesian EN uses the following
two steps to move from $(\beta, \sigma^2)$ to $(\beta', {\sigma^2}')$.
\begin{algorithm}[H]
    \caption{One iteration of the DA algorithm for Bayesian elastic net}
    \begin{algorithmic}[1]
          \STATE Given $(\beta, \sigma^2)$, draw $y_1,\dots, y_p$ independently with
          \[
            \frac{1}{y_j} \Big| \beta, \sigma^2, z\sim \mbox{Inverse-Gaussian} \bigg(\sqrt{\frac{\lambda_1^2 \sigma^2}{\beta^2_j}}, \lambda_1^2\bigg)\;\mbox{for}\; j=1,2,\dots,p.
          \]

                        \STATE Draw $(\beta^\prime, {\sigma^2}^\prime)$ by first drawing
                        \[
                          {\sigma^2}^\prime|y, z \sim \mbox{Inverse-Gamma} \Big( \frac{n-1}{2} + \alpha, \frac{\tilde{z}^\top(I - W(W^{\top}W + \tilde{D}_y^{-1})^{-1}W^{\top})\tilde{z}}{2}+\xi \Big),
                          \]
 and then drawing

              \[
                {\beta}^\prime | {\sigma^2}^\prime, y, z \sim N_p ((W^\top W + \tilde{D}_y^{-1})^{-1}
   W^{\top} \tilde{z}, {\sigma^2}^\prime(W^{\top}W + \tilde{D}_y^{-1})^{-1}).
                \]

	\end{algorithmic}
      \end{algorithm}
      Different Bayesian generalized lasso methods, such as the
      Bayesian group lasso, the Bayesian sparse group lasso, and the
      Bayesian fused lasso models have been proposed in the literature
      to handle the situations where the covariates are known to have
      some structures, for example, when they form groups or are
      ordered in some way \citep{kyun:gill:ghos:case:2010,
        xu:ghos:2015}. Introducing appropriate augmented variables $y$,
      DA algorithms for these models can also be constructed
      \citep{kyun:gill:ghos:case:2010, jin:tan:2021}.
\subsection{Data augmentation for Bayesian logistic models}
\label{sec:logit}
Logistic regression model is likely the most popular model for
analyzing binomial data. Let $(Z_1, Z_2, \dots, Z_m)$ be a
vector of independent Binomial random variables, and $w_i$ be the $p \times 1$ vector
of known covariates associated with $Z_i$ for
$i=1,\dots,m$.  Let $\beta \in \mathbb{R}^p$ be the vector of unknown
regression coefficients. Assume that $Z_i \sim $ Binomial
$(\ell_i, F(w_i^\top \beta))$ where $F(t)= e^t/(1+e^t)$ is the
cumulative distribution function of the standard logistic random
variable. Denoting the observed responses by
$z=(z_1, z_2, \dots, z_m)$, the likelihood function for the logistic
regression model is
\[
L(\beta|z)=\prod_{i=1}^{m} {\ell_i \choose z_i} \frac{\left[\exp\left(w_{i}^\top\beta\right)\right]^{z_{i}}}{[1+\exp\left(w_{i}^\top \beta\right)]^{\ell_i}}.
\]
Consider a Bayesian
analysis that employs the following Gaussian prior for $\beta$
\begin{align}
\label{eq:betaprior}
\pi(\beta) \propto \exp \Big[-\frac{1}{2}(\beta-\mu_{0})^\top Q(\beta-\mu_{0}) \Big],
\end{align}
where $\mu_{0} \in \mathbb{R}^{p}$ and $Q$ is a $p \times p$ positive
definite matrix. Then the intractable posterior density of $\beta$ is given by
\begin{equation}
  \label{eq:logibetapost}
\pi(\beta \,|\, z) \propto \frac{L(\beta|z) \pi(\beta)}{c(z)} = \frac{1}{c(z)}\prod_{i=1}^{m} {\ell_i \choose w_i} \frac{\left[\exp\left(w_{i}^\top\beta\right)\right]^{z_{i}}}{[1+\exp\left(w_{i}^\top \beta\right)]^{\ell_i}}  \pi(\beta) \;,  
\end{equation}
where
\begin{equation}
  \label{eq:logimargden}
  c(z) = \int_{\mathbb{R}^p} \prod_{i=1}^{m} {\ell_i \choose w_i} \frac{\left[\exp\left(w_{i}^\top\beta\right)\right]^{z_{i}}}{[1+\exp\left(w_{i}^\top \beta\right)]^{\ell_i}}  \pi(\beta) d\beta
\end{equation}
is the marginal density of $z$.

There have been many attempts \citep{holm:held:2006, fruh:fruh:2010} to produce an efficient DA algorithm for
the logistic regression model, that is \eqref{eq:logibetapost},
without much success until recently, when
\cite{pols:scot:wind:2013} (denoted as PS\&W hereafter) proposed a new
DA algorithm based on the P{\'o}lya-Gamma (PG)
distribution. A random variable $\theta$ follows a PG
distribution with parameters $a>0, \, b \ge 0$, if
\[
  \theta \overset{d}{=} (1/(2\pi^2)) \sum_{i =1}^{\infty} g_{i} /
  [(i-1/2)^2 +b^2/(4\pi^2)],
\]
where
$g_{i} \overset{iid}{\sim}$Gamma$(a,1)$. From \cite{wang:roy:2018b}, the pdf for PG$(a,b), a>0, \,b \ge 0$ is 
\begin{equation}
 \label{eq:pg1}
p(\theta \mid a,b) = \Big[\cosh \big(\frac{b}{2}\big)\Big]^a\frac{2^{a - 1}}{\Gamma{(a)}} \sum_{r=0}^{\infty}(-1)^r \frac{\Gamma{(r+a)}}{\Gamma{(r+1)}}\frac{(2r+a)}{\sqrt{2\pi \theta^{3}}} \exp\Big(-\frac{(2r+a)^{2}}{8\theta} - \frac{\theta b^{2}}{2}\Big), 
\end{equation}
for $\theta>0,$ where the hyperbolic cosine function $\cosh(t) = (e^{t} + e^{-t})/2$. We denote this as
$\theta \sim $PG$(a,b)$. 

Let $y=(y_{1},y_{2},...y_{n})$, $k_{i} = z_{i} -\ell_i/2, \, i = 1,...,n$ and $p(y_{i}\mid \ell_i,0)$ be the pdf of $y_{i} \sim $PG$(\ell_i,0)$. 
Define the joint posterior density of $\beta$ and $y$ given $z$ as
\begin{equation}
  \label{eq:logijt}
    \pi(\beta, y \mid z) = \frac{1}{c(z)} \bigg[\prod_{i=1}^{m} \frac{ \exp\{k_{i}w_{i}^\top \beta-y_{i}(w_{i}^\top\beta)^{2}/2\}}{2}p(y_{i}|\ell_i, 0)\bigg] \pi(\beta).
  \end{equation}
  By Theorem 1 in \cite{pols:scot:wind:2013}, it follows that the
  $\beta$-marginal density of the joint posterior density
  $\pi(\beta, y \mid z)$ is the target density $\pi(\beta \,|\, z)$
  given in \eqref{eq:logibetapost}, that is,
\begin{align*}
\label{eq:logint}
\pi(\beta \mid z) = \int_{\mathbb{R}_{+}^{m}} \bigg[\prod_{i=1}^{m} \frac{ \exp\{k_{i}w_{i}^\top \beta-y_{i}(w_{i}^\top\beta)^{2}/2\}}{2}p(y_{i}\mid \ell_i,0)\bigg] d y\times \frac{\pi(\beta)}{c(z)}.
\end{align*}
  
PS\&W's DA
algorithm for
the logistic regression model is based on the joint density $\pi(\beta, y \mid z)$. We now derive the
conditional densities, $\pi(\beta| y, z)$ and $\pi(y|\beta, z)$.
From \eqref{eq:logijt} we see that
\begin{equation}
  \label{eq:pg_y}
  \pi(y_{i} \mid \beta, z) \propto \exp(-y_{i}(w_{i}^\top \beta)^{2}/2) p(y_{i} | \ell_i, 0),
\end{equation}  
and thus from \eqref{eq:pg1} we have
$y_{i}|\beta, z \overset{ind}\sim\text{PG}\left(\ell_i,\left|w_i^\top
    \beta\right|\right),$ for $i=1,\dots,m$. Let $W$ denote the
$m\times p$ design matrix with $i$th row $w_i^\top$ and $Y$ be the
$m\times m$ diagonal matrix with $i$th diagonal element $y_i$. From
\eqref{eq:logijt} and \eqref{eq:betaprior}, it follows that the
conditional density of $\beta$ is
\begin{equation*}
\label{eq:cond_beta}
\pi\left(\beta|y, z\right) \propto\exp\left[-\frac{1}{2}\beta^{\top} (W^{\top} Y W +Q)\beta+\beta^{\top}(W^{\top}\kappa + Q\mu_0)\right],
\end{equation*}
where $\kappa=\left(\kappa_{1},\dots,\kappa_{m}\right)^{\top}$. Thus the conditional distribution of $\beta$ is multivariate normal. In particular,
\begin{equation}
\label{eq:normal_beta}
\beta| y, z  \sim N\left(\left(W^{\top}YW +Q \right)^{-1}(W^{\top}\kappa+Q\mu_0),\left(W^{\top}YW+Q\right)^{-1}\right).
\end{equation}
Thus, a single iteration of PS\&W's DA algorithm uses the following
two steps to move from $\beta$ to $\beta'$.

\begin{algorithm}[H]
    \caption{One iteration of the PS\&W's DA algorithm}
    \begin{algorithmic}[1]

          \vspace{.1in}

          \STATE Given $\beta$, draw $y_1,\dots, y_m$ independently with $y_{i} \sim \text{PG}\left(\ell_i,\left|w_i^\top    \beta\right|\right)$.

              \vspace{.1in}

              \STATE Draw $\beta' \sim N\left(\left(W^{\top}YW +Q\right)^{-1}(W^{\top}\kappa +Q\mu_0),\left(W^{\top}YW+Q\right)^{-1}\right)$.

                  \vspace{.1in}

	\end{algorithmic}
\end{algorithm}
PS\&W's DA algorithm is also applicable to the situation when $Q = 0$
in \eqref{eq:betaprior}, that is, $\pi(\beta) \propto 1$, the improper
uniform prior on $\beta$ provided the posterior density $\pi(\beta|z)$
is proper, that is $c(z)$ defined in \eqref{eq:logimargden} is
finite. \cite{pols:scot:wind:2013} demonstrate superior empirical
performance of their DA algorithm over some other DA and MCMC
algorithms for the logistic regression model. 
\subsection{Data augmentation for probit mixed models}
\label{sec:probit}
Generalized linear mixed models (GLMMs) are often used for analyzing
correlated binary observations. The random effects in the linear
predictor of a GLMM can accommodate for overdispersion as well as
dependence among correlated observations arising from longitudinal or
repeated measures studies. Let $(Z_1, Z_2, \dots, Z_m)$ denote the vector of Bernoulli
random variables. Let $w_i$ and $v_i$ be the
$p \times 1$ and $q \times 1$ known covariates and random effect
design vectors, respectively associated with the $i$th observation $Z_i$ for $i=1,\dots,m$.
Let $\beta \in \mathbb{R}^p$ be the unknown vector of regression
coefficients and $u \in \mathbb{R}^q$ be the random effects
vector.  The probit GLMM  connects the expectation of $Z_i$ with 
$w_i$ and $v_i$ using the probit
link function, $\Phi^{-1} (\cdot)$ as
\begin{equation}
  \label{eq:problink}
\Phi^{-1} (P(Z_i =1)) = w_i^\top \beta + v_i^\top u,  
\end{equation}
 where $\Phi(\cdot)$ is the cumulative distribution function of the
standard normal random variable.

Assume that we have $r$ random
effects with $u = (u_1^\top,\dots, u_r^\top)^\top$, where
$u_j$ is a $q_j\times1$ vector with $q_j >0$, $q_1+\cdots + q_r = q$, and
$u_j \stackrel{\text{ind}}{\sim} N(0, \Lambda_j \otimes R_j)$ where the
low-dimensional covariance matrix $\Lambda_j$ is unknown and must be
estimated, and the structured matrix $R_j$ is usually known. Here,
$\otimes$ indicates the Kronecker product.
Denoting $\Lambda = (\Lambda_{1},\dots,\Lambda_{r})$,
the probit GLMM is given by
\begin{eqnarray}
\label{eq:data}
Z_{i} | \beta, u , \Lambda& \overset{\text{ind}}\sim & \text{Bern}(\Phi(x_{i}^{\top}\beta+v_{i}^{\top}u)) \text{ for } i=1,\dots,m \;\text{ with} \nonumber\\
u_{j}|\beta, \Lambda & \overset{\text{ind}}\sim & N(0, \Lambda_j \otimes R_j) , \, j=1,\dots,r .\nonumber
\end{eqnarray}
 Let
$z = (z_1, z_2, \dots, z_m)^\top$ be the observed Bernoulli response variables. Note that,
the likelihood function for $(\beta, \Lambda)$ is
\begin{eqnarray}
  \label{eq:likglmm}
  L(\beta, \Lambda | z) = \int_{\mathbb{R}^{q}} \prod_{i=1}^m\left[\Phi(x_{i}^{\top}\beta+v_{i}^{\top}u)\right]^{z_i}\left[1-\Phi(x_{i}^{\top}\beta+v_{i}^{\top}u)\right]^{1-z_i} \phi_q(u; 0, A(\Lambda))d u,
\end{eqnarray}
which is not available in closed form. Here,
$A(\Lambda) = \oplus_{j=1}^r \Lambda_j \otimes R_j$, with $\oplus$ indicating
the direct sum, and $\phi_q(s; a, B)$ denotes the probability density
function of the $q-$dimensional normal distribution with mean vector
$a$, covariance matrix $B$ and evaluated at $s$.

There are two widely used Monte Carlo approaches for approximating the
likelihood function \eqref{eq:likglmm} and making inference on
$(\beta, \Lambda)$, namely the Monte Carlo EM algorithm
\citep{boot:hobe:1999} and the Monte Carlo maximum likelihood based on
importance sampling \citep{geye:thomp:1992, geye:1994a}. As explained
in \cite{roy:2022}, both the Monte Carlo EM and the Monte Carlo
maximum likelihood methods for making inference on $(\beta, \Lambda)$
require effective methods for sampling from the conditional density
of the random effect $u$
\begin{equation}
  \label{eq:postu}
  f(u|\beta, \Lambda, z) = \frac{f(z, u | \beta,\Lambda)}{L(\beta,\Lambda | z)},
\end{equation}
where $f(z, u | \beta,\Lambda)$ is the joint density of $(z, u)$ given
by
\begin{equation}
  \label{eq:complik}
  f(z, u | \beta,\Lambda ) = \Bigg[\prod_{i=1}^m\left[\Phi(x_{i}^{\top}\beta+v_{i}^{\top}u)\right]^{z_i}\left[1-\Phi(x_{i}^{\top}\beta+v_{i}^{\top}u)\right]^{1-z_i}\Bigg] \phi_{q}(u;0,A(\Lambda)).
\end{equation}
Since the likelihood function
$L(\beta, \Lambda| z)$ is not available in closed form, neither is the
density \eqref{eq:postu}.  


\pcite{albe:chib:1993} DA algorithm for the probit regression model is
one of the most widely used DA algorithms and it can be extended to construct DA samplers for
\eqref{eq:postu}. Following
\cite{albe:chib:1993}, let $y_i \in \mathbb{R}$ be the latent
continuous normal variable corresponding to the $i$th binary observation
$z_i$, that is $z_i = I(y_i >0)$, where
$y_i | \beta, u, \tau \overset{\text{ind}} \sim N(w_i^{\top} \beta +
v_{i}^{\top}u$, 1) for $i =1,\dots, m$. Then
\begin{equation}
  \label{eq:vy}
  P(Z_i=1) = P(Y_i >0) =\Phi(x_i^{\top} \beta +
v_{i}^{\top}u),
\end{equation}
that is,
$Z_{i} | \beta, u, \tau \overset{\text{ind}}\sim
\text{Bern}(\Phi(x_{i}^{\top}\beta+v_{i}^{\top}u))$ as in
\eqref{eq:problink}.  Using the latent variables
$y = (y_1,y_2,\dots,y_m)$, we now introduce the joint density
 \begin{align}
\label{eq:jointprobit}
f(u, y|\beta, \Lambda, z) &= \Bigg[\prod_{i=1}^m \exp\left\{-\frac{1}{2}\left( y_i - w_i^\top \beta - v_i^\top u \right)^2 \right\} \times  \prod_{i=1}^m  \left[1_{\text{(0,\ensuremath{\infty})}}\left(y_{i}\right)\right]^{z_{i}}\left[1_{\left(-\infty,0\right]}\left(y_{i}\right)\right]^{1-z_{i}}\Bigg] \nonumber \\
& \hspace{1in} \times \frac{\phi_q(u; 0, A(\Lambda))}{L(\beta,\Lambda | z)}.
\end{align}
From \eqref{eq:likglmm} and \eqref{eq:vy} it follows that 
\begin{equation}
  \label{eq:margjt}
  \int_{\mathbb{R}^{m}} f(u, y|\beta, \Lambda, z) d y  = f(u|\beta, \Lambda, z),
\end{equation}
that is, the $u-$ marginal of the joint density
$f(u, y|\beta, \Lambda, z)$ is the target density
$f(u|\beta, \Lambda, z)$ given in
\eqref{eq:postu}. Thus, the augmented data $y$ and the joint
density \eqref{eq:jointprobit} satisfy the first property for
constructing a valid DA algorithm for $f(u|\beta, \Lambda, z)$.

If the two conditionals of the joint density \eqref{eq:jointprobit}
are easy to sample from, then a DA algorithm can be constructed. From \eqref{eq:jointprobit}, we see that
\begin{equation}
\label{eq:problatcond}
y_i| u, \beta, \Lambda, z \overset{\text{ind}}\sim \text{TN}(w_i^\top \beta + v_i^\top u,1,z_i), \, i=1,\dots, m,
\end{equation}
where $\text{TN}(\mu, \sigma^2, e)$ denotes the distribution of the
normal random variable with mean $\mu$ and variance $\sigma^2$, that
is truncated to have only positive values if $e = 1$, and only
negative values if $e = 0$. Let $W$ and $V$ denote the $m \times p$
and $m \times q$ matrices whose $i$th rows are $w^\top_i$ and
$v^\top_i$, respectively. Next, using standard linear models-type
calculations, \cite{roy:2022} shows that the conditional distribution of
$u$ is
\begin{equation}
  \label{eq:probucond}
  u|y, \beta, \Lambda, z \sim N_{q}\left( (V^{\top}V + A(\Lambda)^{-1})^{-1} V^{\top} (y - W\beta), (V^{\top}V + A(\Lambda)^{-1})^{-1}\right).
\end{equation}
Thus, a single iteration of the DA algorithm uses the following
two steps to move from $u$ to $u'$. 
\begin{algorithm}[H]
  \caption{One iteration of the DA algorithm for the probit GLMM}
  \label{alg:probda}
  \begin{algorithmic}[1]
              \vspace{.1in}

  \STATE Given $u$, draw $y_i \overset{\text{ind}}\sim \text{TN}(w_i^{\top}\beta+v_i^{\top} u,1,z_i)$ for  $i=1,\dots, m$.
            \vspace{.1in}

            \STATE Draw $u' \sim N_{q}\left( (V^{\top}V + A(\Lambda)^{-1})^{-1} V^{\top} (y - W\beta), (V^{\top}V + A(\Lambda)^{-1})^{-1}\right)$.
                      \vspace{.1in}

\end{algorithmic}
\end{algorithm}

One can consider a Bayesian analysis of the probit GLMM with
appropriate priors on $(\beta, \Lambda)$ and the DA and Haar PX-DA
algorithms discussed in this section can be extended to sample from
the corresponding posterior densities
\cite[see][]{wang:roy:2018,roy:2022}. Similarly, PS\&W's DA
algorithm for the logistic model discussed in Section~\ref{sec:logit}
can be extended to fit logistic mixed models
\cite[see][]{wang:roy:2018a,rao:roy:2021,roy:2022}.



\section{Improving the DA algorithm}
\label{sec:imprda}
DA algorithms, although popular MCMC schemes, can suffer from
slow convergence to stationarity \citep{roy:hobe:2007,
  hobe:roy:robe:2011, duan:john:2018}. In the literature, a lot of
effort has gone into developing methods for accelerating the speed of
convergence of the DA algorithms. These methods are mainly based on
techniques like appropriate reparameterizations and parameter
expansions such as the centered and noncentered parameterizations
\citep{papa:robe:2008}, interweaving the two parameterizations
\citep{yu:meng:2011}, {\it parameter expanded-data augmentation}
(PX-DA) \citep{liu:wu:1999}, the {\it marginal augmentation}
\citep{meng:vand:1999} or the calibrated DA \citep{duan:john:2018}.
Unlike the other methods mentioned here, \pcite{duan:john:2018} calibrated DA
requires an additional Metropolis-Hastings (MH) step to maintain the
stationarity of the algorithm with respect to the target density
$f_X$. The PX-DA and marginal augmentation are among the most popular
techniques for speeding up DA algorithms, and we describe those in
more details later in this section.


Graphically, the two steps of the DA algorithm can be viewed as
$x \rightarrow y \rightarrow x'$. It has been observed that the
convergence behavior of a DA algorithm can be significantly improved
by inserting a properly chosen extra step in between the two steps of
the DA algorithm. The idea of introducing an extra step using
appropriate auxiliary variables was developed independently by
\cite{liu:wu:1999}, who called the resulting MCMC algorithm the PX-DA,
and \cite{meng:vand:1999}, who called it the marginal
augmentation. Following these works, \cite{hobe:marc:2008} developed
general versions of PX-DA type algorithms and \cite{yu:meng:2011}
referred to these generalized methods as {\it sandwich algorithms} since the algorithms
involve an additional step that is sandwiched between the two steps of
the original DA algorithm. Graphically, a sandwich algorithm can be
represented as $x \rightarrow y \rightarrow y' \rightarrow x'$, where
the first and the third steps are the two steps of the DA algorithm.
Suppose the middle step $y \rightarrow y'$ is implemented by making a
draw according to a Mtd $r(y'|y)$. Thus, denoting the current state by
$x$, a generic sandwich algorithm uses the following three steps to
move to the new state $x'$.
\begin{algorithm}[H]
  \caption{An iteration of a generic sandwich algorithm}
  \label{alg:sand}
  \begin{algorithmic}[1]

    \vspace{.1in}
    
    \STATE Given $x$, draw $Y \sim f_{Y|X}(\cdot|x)$, and call the observed value $y$.
\vspace{.1in}
    
    \STATE Draw $Y' \sim r(y'|y)$, and call the result $y'$.
    \vspace{.1in}
    
    \STATE Draw $X' \sim f_{X|Y}(\cdot|y')$.
    \vspace{.1in}
    
\end{algorithmic}
\end{algorithm}

We now describe how to construct an Mtd $r$ for a valid middle step
$y \rightarrow y'$ for sandwich algorithms.  Recall that $f_Y(y)$ is
the marginal density of $y$. Suppose $r(y'|y)$ leaves $f_Y$ invariant, that is,
\begin{equation}
   \label{eq:invar}
  \int_{\mathbb{R}^q} r(y'|y) f_Y(y) \, dy = f_Y(y') \;.
\end{equation}
It turns out that {\it any} $r$ satisfying this invariance condition
implies that $f_X$ is stationary for the corresponding sandwich
algorithm and consequently can be used to obtain approximate samples
from $f_X$. To see it, note that the Mtd of the sandwich algorithm is
given by
 \begin{equation}
  \label{eq:mtdsa}
  k_{\text{SA}}(x'|x) = \int_{\mathbb{R}^q} \int_{\mathbb{R}^q} f_{X|Y}(x'|y') \, r(y'|y) \,
  f_{Y|X}(y|x) \, dy \, dy' \;,
\end{equation}
and
\begin{align}
  \label{eq:invsa}
  \int_{\mathbb{R}^p} k_{\text{SA}}(x'|x) f_X(x) \, dx & = \int_{\mathbb{R}^p} \bigg[ \int_{\mathbb{R}^q}
  \int_{\mathbb{R}^q} f_{X|Y}(x'|y') \, r(y'|y) \, f_{Y|X}(y|x) \, dy \, dy'
  \bigg] f_X(x) \, dx \nonumber \\ & = \int_{\mathbb{R}^q} f_{X|Y}(x'|y') \bigg[
  \int_{\mathbb{R}^q} r(y'|y) \, f_Y(y) \, dy \bigg] \, dy' \nonumber \\ & =
  \int_{\mathbb{R}^q} f_{X|Y}(x'|y') \, f_Y(y') \, dy' \nonumber \\ & = f_X(x'),
\end{align}
where the third equality follows from \eqref{eq:invar}. Furthermore,
if $r(y'|y)$ satisfies the detailed balance condition with respect to
$f_Y(y)$, that is, $r(y'|y) f_Y(y)$ is symmetric in $(y, y')$, then
\begin{align}
    \label{eq:revsa}
    k_{\text{SA}}(x'|x) f_X(x) & = f_X(x) \int_{\mathbb{R}^q} \int_{\mathbb{R}^q} f_{X|Y}(x'|y') \,
    r(y'|y) \, f_{Y|X}(y|x) \, dy \, dy' \nonumber\\ & = \int_{\mathbb{R}^q} \int_{\mathbb{R}^q}
    f_{X|Y}(x'|y') \, r(y'|y) \, f(x,y) \, dy \, dy' \nonumber\\ & = \int_{\mathbb{R}^q}
    \int_{\mathbb{R}^q} f_{X|Y}(x'|y') \, r(y'|y) f_Y(y) \, f_{X|Y}(x|y) \, \, dy
    \, dy' \nonumber\\ & = \int_{\mathbb{R}^q} \int_{\mathbb{R}^q} f_{X|Y}(x'|y') \, r(y|y') f_Y(y') \,
    f_{X|Y}(x|y) \, \, dy \, dy' \nonumber\\ & = \int_{\mathbb{R}^q} \int_{\mathbb{R}^q} f(x',y') \,
    r(y|y') \, f_{X|Y}(x|y) \, \, dy \, dy' \nonumber\\ & = f_X(x') \int_{\mathbb{R}^q}
    \int_{\mathbb{R}^q} f_{X|Y}(x|y) \, r(y|y') \, f_{Y|X}(y'|x') \, \, dy \, dy'\nonumber
    \\ & = k_{\text{SA}}(x|x') f_X(x') \;,
\end{align}
that is, the corresponding sandwich algorithm is reversible
with respect to the target density $f_X$.

Let us denote the Markov chain underlying the sandwich algorithm by
$\{X_n^*\}_{n \ge 1}$. Since from \eqref{eq:invsa} or \eqref{eq:revsa}
it follows that the target density $f_X$ is stationary for
$\{X_n^*\}_{n \ge 1}$, the sample averages
$\bar{h}^*_n := \sum_{i=1}^n h(X^*_i)/n$ based on the SA Markov chain
can also be used to estimate $E_{f_X} h$. Indeed, as mentioned in Section~\ref{sec:intro}, if
$\{X_n^*\}_{n \ge 1}$ is appropriately irreducible,
$\bar{h}^*_n \rightarrow \mbox{E}_{f_X}h$ almost surely as
$n \rightarrow \infty$. Now, $\bar{h}^*_n$ will be preferred over
$\bar{h}_n$ as an estimate of $\mbox{E}_{f_X}h$ if the former leads to
smaller (Monte Carlo) errors. In particular, if there is a Markov
chain central limit theorem (CLT) for $\bar{h}_n$, that is, if
$\sqrt{n}(\overline{h}_n - E_{f_X} h) \stackrel{d}{\rightarrow}
\mathcal{N}(0, \sigma_h^2)$ for some finite asymptotic variance
$\sigma_h^2$, the standard error of $\overline{h}_n$ is
$\hat{\sigma}_h/\sqrt{n}$ where $\hat{\sigma}_h$ is it consistent
estimator of $\sigma_h$ \cite[see][for methods of constructing
$\hat{\sigma}_h$]{fleg:jone:2010, vats:fleg:jone:2019}. Thus, in
practice, establishing CLTs for $\overline{h}_n$ and
$\overline{h}^*_n$ is important for ascertaining and comparing the
quality of these estimators. A sufficient condition extensively used
in the literature for guaranteeing the existence of such a CLT for
every square integrable function $h$
($\int_{\mathbb{R}^p} h^2(x) \, f_X(x) \, dx < \infty$) is that the
Markov chain converges to its stationary distribution at a geometric
rate \citep{robe:rose:1997,chan:geye:1994}.

Formally, the chain
$\{X_n\}_{n \ge 1}$ is called \textit{geometrically ergodic} if there
exist a function $M: \mathbb{R}^p \rightarrow [0,\infty)$ and a constant
$\rho \in [0,1)$ such that, for all $x \in \mathbb{R}^p$ and all
$n=1,2,\dots$,
\begin{equation*}
  \label{eq:ge}
  \int_{\mathbb{R}^p}|k^n(x'|x) - f_X(x)| dx \le M(x) \, \rho^n \;,  
\end{equation*}
where $k^n(\cdot| \cdot)$ is the $n-$step Mtd of the chain
$\{X_n\}_{n \ge 1}$.  Unfortunately, the simple irreducibility and
aperiodicity conditions mentioned in Section~\ref{sec:intro} does not
imply geometric ergodicity.  The most common method of proving
geometric ergodicity of $\{X_n\}_{n \ge 1}$ is by establishing 
\textit{drift and minorization conditions} \citep{rose:1995,
  jone:hobe:2001}. Indeed, several DA algorithm examples considered in
this chapter have been shown to be geometrically ergodic by
establishing appropriate drift and minorization conditions. 
For example, for the Bayesian lasso model discussed in
Section~\ref{sec:bvs}, \cite{raja:spar:khar:zhan:2019} established
geometric ergodicity of the DA Markov chain in the special case when
$\alpha=\xi=0$. 
The geometric
rate of convergence of PS\&W's DA Markov chain for the binary logistic
regression model discussed in Section~\ref{sec:logit} has been established by \cite{choi:hobe:2013} and
\cite{wang:roy:2018b} under proper normal and improper uniform prior on $\beta$, respectively. 
      
If the SA Markov chain $\{X_n^*\}_{n \ge 1}$ is reversible, denoting
the asymptotic variance for $\overline{h}^*_n$ by $\sigma_{h,*}^2$, it
turns out that $\sigma_{h,*}^2 \leq \sigma_h^2$ for every square
integrable (with respect to $f_X$) function $h$
\citep{hobe:marc:2008}. Thus, sandwich algorithms are asymptotically
more efficient than the corresponding DA algorithms.  So if both the
DA and sandwich algorithms are run for the same number of iterations,
then the errors in $\bar{h}^*_n$ are expected to be smaller than that
of $\bar{h}_n$. In other words, to achieve the same level of
precision, the sandwich algorithm is expected to require fewer
iterations than the DA algorithm. In Section~\ref{sec:convintro}, we
discuss other results established in the literature showing the
superiority of the sandwich algorithms over the corresponding DA
algorithms.

Intuitively, the extra step $R$ reduces the correlation between $x$
and $x'$ and thus improves the {\it mixing} of the DA algorithm. On
the other hand, since the extra step in a sandwich algorithm involves
more computational effort compared to the DA algorithm, any gain in
mixing should be weighed against this increased computational burden.
Fortunately, there are several examples where the sandwich algorithm
provides a ``free lunch'' in that it converges much faster than the
underlying DA algorithm while requiring a similar amount of
computational effort per iteration \citep[see
e.g.][]{laha:dutt:roy:2017, pal:khare:2015, roy:hobe:2007,
  hobe:roy:robe:2011, roy:2014}. Following \cite{liu:wu:1999} and
\cite{meng:vand:1999}, the Mtd $r$ for these efficient sandwich
algorithms are often constructed by introducing extra parameters
$g \in \G \subset \mathbb{R}^d$ and a class of functions
$t_g: \mathbb{R}^q \rightarrow \mathbb{R}^q$ indexed by $g$. The
middle step $y \rightarrow y'$ is implemented by drawing $g$ from a
density depending on $y$ and then setting $y' = t_g(y)$ maintaining
invariance with respect to $f_Y$. Typically, $d$ is small, say 1 or 2,
and $\{t_g(y) : g \in G \}$ is a small subspace of
$\mathbb{R}^q$. Thus, drawing from such an $r$ is usually much less
computationally expensive than the two steps of the DA Algorithm. However, as mentioned before, even when
$d=1$, this perturbation $y \rightarrow y'$ often greatly improves the
mixing of the DA algorithm.

We now describe \pcite{liu:wu:1999} Haar PX-DA algorithm where the set
$\G$ is assumed to possess a certain group structure \cite[see
also][]{hobe:marc:2008}. In particular, $\G$ is assumed to be a
topological group; that is, a group such that the functions
$(g_1,g_2) \mapsto g_1 g_2$ and the inverse map $g \mapsto g^{-1}$ are
continuous. An example of such a $\G$ is the \textit{multiplicative group},
$\mathbb{R}_+$, where the binary operation defining the group is
multiplication, the identity element is 1, and $g^{-1} =
1/g$. Suppose, $t_g(y)$ represents $\G$ acting topologically on the
left of $\mathbb{R}^q$ \citep[][Chapter 2]{eato:1989}, that is,
$t_e(y) = y$ for all $y \in \mathbb{R}^q$ where $e$ is the identity
element in $\G$ and $t_{g_1 g_2}(y) = t_{g_1} \big( t_{g_2}(y) \big)$
for all $g_1,g_2 \in \G$ and all $y \in \mathbb{R}^q$.

We assume the existence of a \textit{multiplier}
$\chi : \G \rightarrow \mathbb{R}_+$, that is, $\chi$ is continuous and
$\chi(g_1 g_2) = \chi(g_1) \chi(g_2)$ for all $g_1,g_2 \in \G$.
Assume that the Lebesgue measure on $\mathbb{R}^q$ is \textit{relatively (left)
  invariant} with respect to the multiplier $\chi$; that is, assume that for any
$g \in \G$ and any integrable function $\zeta: \mathbb{R}^q \rightarrow \mathbb{R}$,
we have
\[
\chi(g) \int_{\mathbb{R}^q} \zeta \big( t_g(y) \big) \, dy = \int_{\mathbb{R}^q} \zeta(y) \, dy \;.
\]
Next, suppose the group $\G$ has a \textit{left-Haar measure} of the
form $\nu_l(g) \, dg$ where $dg$ denotes Lebesgue measure on $\G$. It
is known that the left-Haar measure satisfies
\begin{equation}
  \label{eq:haar_def}
  \int_\G h \big( \tilde{g} g \big) \, \nu_l(g) \, dg = \int_\G h(g) \,
  \nu_l(g) \, dg  
\end{equation}
for all $\tilde{g} \in \G$ and all integrable functions
$h : \G \rightarrow \mathbb{R}$.  Finally, assume that
\[
q(y) := \int_\G f_Y \big(t_g(y) \big) \, \chi(g) \, \nu_l(g) \, dg
\]
is strictly positive for all $y \in \mathbb{R}^q$ and finite for (almost) all $y
\in \mathbb{R}^q$.  

We now state how the middle step $y \rightarrow y'$ is implemented in
\pcite{liu:wu:1999} Haar PX-DA algorithm.  If the current state of the
Haar PX-DA chain is $x$, an iteration of this chain uses the following
steps to move to $x'$.

\begin{algorithm}[H]
  \caption{An iteration for the Haar PX-DA Algorithm}
  \label{alg:pxda}
  \begin{algorithmic}[1]

    \vspace{.1in}
    
    \STATE Given $x$, draw $Y \sim f_{Y|X}(\cdot|x)$, and call the observed value $y$.

    \vspace{.1in}
    
    \STATE Draw $G$ from the density proportional to $f_Y \big(t_g(y) \big)
  \, \chi(g) \, \nu_l(g)$, call the result $g$, and set $y' = t_g(y)$.

        \vspace{.1in}
    
    \STATE Draw $X' \sim f_{X|Y}(\cdot|y')$.
    \vspace{.1in}
    
\end{algorithmic}
\end{algorithm}
\begin{example}[continues=sec:probit]
Improving the DA algorithm for the probit mixed model presented in Section~\ref{sec:probit}, \cite{roy:2022} constructs a Haar
PX-DA algorithm for \eqref{eq:postu}. 
For constructing the sandwich step
$y \rightarrow y'$, we need the marginal density of $Y$,
$f_Y(y|\beta, \Lambda, z)$ from the joint density
$f(u, y|\beta, \Lambda, z)$ given in \eqref{eq:jointprobit}. It turns out that
\begin{eqnarray}
\label{eq:margthet}
 f_Y(y|\beta, \Lambda, z) \propto \prod_{i=1}^{m}\left[1_{\text{(0,\ensuremath{\infty})}}\left(y_{i}\right)\right]^{z_{i}}\left[1_{\left(-\infty,0\right]}\left(y_{i}\right)\right]^{1-z_{i}} \exp\left\{ -\frac{1}{2}\left[y^{\top}V_1y-2y^{\top}V_1W\beta\right]\right\},
\end{eqnarray}
where
\[
V_1 = \left[I_m-V\left(V^\top V + (A(\Lambda))^{-1}\right)^{-1}V^{\top}\right].
\]
 Let $\mathcal{Y}$ denote the subset of $\mathbb{R}^m$ where $y$
lives, that is, $\mathcal{Y}$ is the Cartesian product of $m$ half
(positive or negative) lines, where the $i$th component is
$(0, \infty)$ (if $z_i =1$) or $(-\infty, 0]$ (if $z_i =0$).

As mentioned in this Section, 
take $\G = \mathbb{R}_+$ and
let the multiplicative group $\G$ act on $\mathcal{Y}$ through
$t_g(y) = gy = (gy_1,\dots,gy_m)$. Note
that, for any $\tilde{g} \in \G$, we have
\[
\int_0^\infty h \big( \tilde{g} g \big) \, \frac{1}{g} \, dg =
\int_0^\infty h(g) \, \frac{1}{g} \, dg \;,
\]
which shows from \eqref{eq:haar_def} that $\nu(dg)=dg/g$ is a left-Haar measure for the
multiplicative group, where $dg$ is Lebesgue measure on
$\mathbb{R}_+$. Also, with the group action defined this way, it is known that the Lebesgue measure on $\mathcal{Y}$ is
relatively left invariant with the multiplier $\chi (g) = g^m$
\citep{hobe:marc:2008}. Thus,
\begin{equation}
  \label{eq:gprob}
  f_Y\left(gy|\beta, \Lambda, z\right) \chi \left( g\right) \nu(dg) \propto  g^{m-1}\exp\left\{ -\frac{1}{2}\left[g^{2}y^{\top} V_1 y -  2g y^{\top} V_1 W\beta \right]\right\} dg,  
\end{equation}
and since $V_1$ is a positive definite matrix,
\[
  q(y) = \int_0^{\infty} g^{m-1}\exp\left\{ -\frac{1}{2}\left[g^{2}y^{\top} V_1 y -  2g y^{\top} V_1 W\beta \right]\right\} dg
  \]
  is strictly positive for all $y \in \mathcal{Y}$ and finite for
  (almost) all $y \in \mathcal{Y}$. Thus, a single iteration of the Haar PX-DA
  algorithm for the probit mixed model uses the following three steps to move from $u$ to $u'$.
\begin{algorithm}[H]
\caption{One iteration of the Haar PX-DA algorithm for the probit GLMM}
\label{alg:probpxda}
\begin{algorithmic}[1]
  \STATE  Given $u$, draw $y_i \overset{\text{ind}}\sim \text{TN}(w_i^{\top}\beta+v_i^{\top} u,1,z_i)$ for  $i=1,\dots, m$.
  
\STATE Draw $g$ from \eqref{eq:gprob}.

\STATE Calculate
  $y_i^{\prime} = gy_i$ for $i=1,\dots,m$, and draw $u'$ from \eqref{eq:probucond} conditional on
  $y^{\prime} = (y_1^{\prime},\dots, y_m^{\prime})^{\top}$, that is, draw
\[u' \sim N_{q}\left( (V^{\top}V + (A(\Lambda))^{-1})^{-1} V^{\top} (y' - W\beta), (V^{\top}V + A(\Lambda)^{-1})^{-1}\right).\]
 \end{algorithmic}
\end{algorithm}
The density \eqref{eq:gprob} is log-concave, and \cite{roy:2022}
suggests using \pcite{gilk:wild:1992} adaptive rejection sampling
algorithm to sample from it. Since a single draw from \eqref{eq:gprob}
is the only difference between each iteartion of the DA and the Haar
PX-DA algorithms, the computational burden for the two algorithms is similar.  
\end{example}

Given the group $\G$, the Haar PX-DA algorithm is the best sandwich
algorithm in terms of efficiency and operator norm
\citep{hobe:marc:2008}. \cite{khar:hobe:2011} established necessary
and sufficient conditions for the Haar PX-DA algorithm to be strictly
better than the corresponding DA algorithm (see
Section~\ref{sec:convintro} for details). 
Also, the Haar PX-DA method
with appropriate choices of $\G$ and $t_g$ has been empirically
demonstrated to result in a huge improvement in the mixing of DA algorithms
albeit with roughly the same computational cost in various examples
\citep[see e.g.][]{roy:hobe:2007, liu:wu:1999,meng:vand:1999,
  roy:2014, pal:khare:2015}. 
Finally, there are other strategies proposed in the literature to improve DA
algorithms without introducing extra auxiliary variables as in the sandwich algorithms. For example,
\cite{roy:2016} replaces one of the two original steps of
the DA algorithm with a draw from the MH step given in
\cite{liu:1996}. One advantage of \pcite{roy:2016} modified DA algorithm over the sandwich algorithms is that the Markov transition matrix of the modified algorithm can have negative eigenvalues, whereas the DA algorithms and also, generally, the sandwich algorithms used in practice are positive
Markov chains leading to positive eigenvalues
\citep{khar:hobe:2011}. Hence, \pcite{roy:2016} modified DA algorithm
can lead to superior performance over the DA and sandwich algorithms
in terms of asymptotic variance.



 \section{Spectral properties of the DA Markov chain}
 \label{sec:convintro}



The convergence of the Markov chains associated with MCMC algorithms to the desired stationary distribution is at the heart of the popularity of MCMC technology. In Section~\ref{sec:intro} we discussed the convergence of the sample averages $\bar{h}_n$ and the marginal distributions of $X_n$.  
For a serious practitioner, understanding the quality of both the convergences is important. As shown in Section~\ref{sec:imprda}, proving geometric ergodicity of the Markov chain is key for providing standard errors for $\bar{h}_n$. 
Establishing geometric ergodicity for general state space Markov chains arising in statistical practice is in general quite challenging.
The drift and minorization technique that is often used to prove geometric ergodicity has two flavors/tracks.
The first track leads to establishing qualitative geometric ergodicity, without providing quantitative convergence bounds for distance to stationarity after finitely many Markov chain iterations. The other track leads to explicit convergence bounds, but these bounds are often too conservative to be useful, especially in modern high-dimensional settings \cite{jone:hobe:2004, qin:hobe:2021}. While several alternate and promising methods for obtaining convergence bounds have been developed in recent years (see \cite{qin:hobe:2022} for a useful summary), it is safe to say that a successful geometric ergodicity analysis remains elusive for an overwhelming majority of MCMC algorithms used in statistical practice. However, as we will see below, the special structure/construction of the DA Markov chain often allows us in many examples to establish stronger properties which imply geometric ergodicity in 
particular, and lead to a richer and deeper understanding of the structure and convergence properties of the relevant DA Markov chain. 

\subsection{Leveraging operator theory for comparing DA and sandwich chains}

Operator theory and associated tools play a key role in convergence analysis and comparison of the DA Markov chain (with Mtd $k$) and the sandwich Markov chain (with Mtd $k_{SA}$). Let $L_0^2 (f_X)$ denote the space of real-valued functions (with domain $\mathbb{R}^p$) that have mean zero and are square integrable with respect to $f_X$. In particular, 
$$
L_0^2 (f_X) = \left\{ h: \mathbb{R}^p \rightarrow \mathbb{R}: \; \int_{\mathbb{R}^p} h(x)^2 f_X (x) dx < \infty \mbox{ and} \int_{\mathbb{R}^p} h(x) f_X (x) dx = 0 \right\}. 
$$

\noindent
Let $K: L_0^2 (f_X) \rightarrow L_0^2 (f_X)$ and $K_{SA}: L_0^2 (f_X) \rightarrow L_0^2 (f_X)$ denote the {\it Markov operators} corresponding to the Markov transition densities 
$k$ and $k_{SA}$ respectively. Then for any $h \in L_0^2 (f_X)$, we have 
$$
(Kh)(x) = \int_{\mathbb{R}^p} h(x') k(x' \mid x) dx' \mbox{ and } (K_{SA}h)(x) = \int_{\mathbb{R}^p} h(x') k_{SA} (x' \mid x) dx'. 
$$

Note that $L_0^2 (f_X)$ is a Hilbert space equipped with the inner product 
$$
<h_1, h_2> = \int_{\mathbb{R}^p} h_1(x) h_2(x) f_X (x) dx
$$

for every pair $h_1,h_2 \in L_0^2 (f_X)$, and the corresponding norm defined by $\|h\| = 
\sqrt{<h,h>}$ for every $h \in L_0^2 (f_X)$. The operator norms of $K$ and $K_{SA}$, denoted respectively by $\|K\|$ and $\|K_{SA}\|$, are defined as 
$$
\|K\| = \sup_{h \in L_0^2 (f_X), \|h\| = 1} \|Kh\| \mbox{ and } \|K_{SA}\| = \sup_{h \in L_0^2 (f_X), \|h\| = 1} \|K_{SA}h\|. 
$$

The convergence properties of the DA and sandwich Markov chains are intricately linked to the behavior of the corresponding Markov operators $K$ and $K_{SA}$. 
Using the structure of $k$ and $k_{SA}$, it can be shown easily that the operator $K$ is a non-negative operator (an operator $P$ is non-
negative if $<Pg,g>$ is non-negative $\forall g$), $\|K\| \in [0,1]$ and $\|K_{SA}\| \in [0,1]$. Also, $K$ is a self-adjoint operator, and if $r(y' \mid y)$ satisfies detailed balance with respect to $f_Y (y)$ (which will be assumed henceforth, unless specified otherwise) then so is $K_{SA}$ (an operator $P$ is self-adjoint if $<Ph_1,h_2> = <Ph_2,h_1> \; \forall h_1,h_2$). \cite{robe:rose:1997} show that for a self-adjoint Markov operator $P$ corresponding to a Harris ergodic chain, $\|P\| < 1$ if and only if the corresponding Markov chain is geometrically ergodic. In this case, $\|P\|$ in fact corresponds to the asymptotic rate of convergence of the relevant Markov chain to its stationary distribution. The above concepts enable us to state the first result that rigorously shows that the sandwich chain is at least as good as the DA chain in some key aspects. 
\begin{theorem}[\cite{hobe:marc:2008}] \label{thm:rate}
If the DA chain with Mtd $k$ is Harris ergodic, and the sandwich chain (with Mtd $k_{SA}$) can itself be interpreted as a DA chain, then $\|K_{SA}\| \leq \|K\|$. Hence, geometric Harris ergodicity of the DA chain implies geometric Harris ergodicity of the sandwich chain, and in such settings the asymptotic rate of convergence to stationarity for the sandwich chain is at least as good as that of the DA chain.  
\end{theorem}

The structure and analysis of sandwich chains is more complicated (sometimes significantly so) than the DA chains due to the additional sandwich step. However, the above result allows us to completely avoid analyzing the sandwich chain, if geometric ergodicity can be established for the corresponding DA chain. 

Intuition says that given the extra step, the sandwich chain should have better mixing and faster convergence than the original DA chain under suitable 
regularity conditions. The above result provides rigorous justification to this intuition. See Section \ref{trace:comparison} for further discussion. 

\subsection{The trace class property for Markov operators}

An operator $P$ on $L^2_0 (f_X)$ is defined to be {\it Hilbert-Schmidt} if for any orthonormal sequence $\{h_n\}_{n \geq 1}$ in $L^2_0 (f_X)$, we have 
\begin{equation} \label{defhs}
\sum_{n=1}^\infty \|Ph_n\|^2 < \infty. 
\end{equation}

\noindent
Additionally, a positive self-adjoint operator (such as the DA operator $K$) is defined to be {\it trace class} if 
\begin{equation} \label{deftc}
\sum_{n=1}^\infty <Ph_n, h_n> < \infty 
\end{equation}

for any orthonormal sequence $\{h_n\}_{n \geq 1}$ in $L^2_0 (f_X)$. Both the above properties imply that the operator $P$ is compact, and has countably many singular values. The trace class property implies that these singular values are summable, while the Hilbert-Schmidt property implies that these singular values are square-summable. If the associated Markov chain is Harris ergodic, then compactness of a Markov operator implies that its spectral radius is less than $1$. Existing results (see for example Proposition 2.1 and Remark 2.1 in \cite{robe:rose:1997}) can now be leveraged to establish geometric ergodicity. To summarize, for Harris ergodic Markov chains, we have 
$$
\mbox{Hilbert Schmidt } \Longrightarrow \mbox{Compact } \Longrightarrow \mbox{Geometrically ergodic},
$$

\noindent
and if the corresponding Markov operator is also positive and self-adjoint, then 
$$
\mbox{Trace Class } \Longrightarrow \mbox{Hilbert Schmidt } \Longrightarrow \mbox{Compact } \Longrightarrow \mbox{Geometrically ergodic}. 
$$

The above implications potentially offer an alternate mechanism/route for establishing geometric ergodicity. However, the conditions in (\ref{defhs}) and 
(\ref{deftc}) seem hard to verify especially for intractable Markov transition densities arising in statistical applications. Fortunately, there are equivalent characterizations for (\ref{defhs}) and (\ref{deftc}) that are easier to state and verify. In particular, results in \cite{Jorgens:1982} can be leveraged to show that a Markov operator $P$ (with corresponding Mtd $p(\cdot, \cdot)$) on $L^2_0 (f_X)$ is Hilbert-Schmidt if and only if 
\begin{equation} \label{integralhs}
\int_{\mathbb{R}^p} \int_{\mathbb{R}^p} \frac{p(x,y)^2 f_X (x)}{f_X (y)} dx dy = \int_{\mathbb{R}^p} \int_{\mathbb{R}^p} \left( \frac{p(x,y)}{f_X (y)} \right)^2 f_X (x) f_X (y) dx dy < \infty, 
\end{equation}
and in case $P$ is self-adjoint and positive, it is trace class if and only if 
\begin{equation} \label{integraltc}
\int_{\mathbb{R}^p} p(x,x) dx < \infty. 
\end{equation}

\noindent
The conditions in (\ref{integralhs}) and (\ref{integraltc}) while simpler than those in (\ref{defhs}) and (\ref{deftc}) can still be quite challenging to 
verify for Markov chains arising in statistical practice. The corresponding Markov transition densities are themselves often expressed as intractable 
integrals, and the analysis of the expressions in (\ref{integralhs}) and (\ref{integraltc}) can be quite lengthy, laborious and intricate. As the authors state in \cite{liu:wong:kong:1995}, the condition in (\ref{integralhs}) ``is standard, but not easy to check and understand". Yet, there have been 
many success stories in recent years, see \cite{khar:hobe:2011,PKH:2017,chak:khar:2017,Chakraborty:Khare:2019,QHK:2019,raja:spar:khar:zhan:2019,jin:tan:2021}. Markov chains used by statistical practitioners overwhelmingly employ either the Gibbs sampling algorithm, 
or the Metropolis-Hastings algorithm, or a combination of both. Any continuous state space chain with a non-trivial Metropolis component cannot be compact \citep{chan:geye:1994} and 
hence cannot be trace class. The same is true for random scan Gibbs chains. (Systematic scan) Gibbs sampling Markov chains are in general not reversible. 
However, the convergence of a two-block (systematic scan) Gibbs chain is completely characterized by its ``marginal chains" which correspond to positive self-
adjoint Markov operators \citep{liu:wong:kong:1994, diac:khar:salo:2008}. In particular, the construction in Section \ref{sec:intro} shows that the DA chain is 
a marginal chain of a two-block Gibbs sampler for $f_{X,Y}$. Hence, it is not surprising that all Markov chains which have been shown to be trace class in 
these papers are DA Markov chains. 

\subsection{Trace class property as a tool to establish geometric ergodicity}

We have already discussed how establishing the trace class property for a positive self-adjoint Markov operator corresponding to a Harris ergodic Markov chain 
establishes geometric ergodicity of that chain. In this section, we compare the potential advantages and disadvantages of this alternative approach compared to 
the drift and minorization approach for establishing geometric ergodicity. A key advantage is that the trace class approach based on (\ref{integraltc}) is 
more straightforward and streamlined than the drift and minorization approach. Indeed, construction of effective drift functions depends heavily on the Markov 
chain at hand and has been described as ``a matter of art" (\cite{diac:khar:salo:2008}). On the other hand, the drift and minorization approach has broader 
applicability, and has been successfully applied to non-reversible and non-compact chains. 

Suppose that for a DA Markov chain one is able to establish geometric ergodicity through a drift and minorization analysis, and is also able to establish the 
trace class condition in (\ref{integraltc}). Since the trace class property is much stronger than geometric ergodicity, one would expect that assumptions 
needed for establishing the trace class property are stronger than those needed for the drift and minorization analysis. Surprisingly, as demonstrated in \cite{mukh:khar:2023}, this is not always true. 
To understand why this might happen, we take a brief but careful look at the Markov chain analyzed in \cite{mukh:khar:2023} below. 

Similar to Sections \ref{sec:logit} and \ref{sec:probit}, consider a binary regression setting with $n$ independent binary responses $Z_1, Z_2, \cdots, Z_n$ 
and corresponding predictor vectors ${\bf w}_1, {\bf w}_2, \cdots, {\bf w}_n \in \mathbb{R}^p$, such that 
\begin{equation} \label{robitmodel}
P(Z_i = 1 \mid \bb) = F_\nu ({\bf w}_i^T \bb), 
\end{equation}

\noindent
for $1 \leq i \leq n$. The goal is to estimate $\bb \in \mathbb{R}^p$. Here $F_\nu$ is the CDF of the Student's $t$-distribution with $\nu$ degrees of freedom, and the corresponding model is referred to as the robit regression model \cite{liu:2004}. In the binary regression setting, outliers are observations with unexpectedly large predictor values and a misclassified response. The robit model can more effectively down-weight outliers and produce a better fit than probit or logistic models \cite{preg:1982}.

Following \cite{Roy:2012:robit, albe:chib:1993}, consider a Bayesian model with a multivariate normal prior for $\bb$ with mean $\bb_a$ and covariance matrix $\Sigma_a^{-1}$. Let ${\bf z} \in \{0,1\}^n$ denote the vector of observed values for the response variables $Z_1, Z_2, \cdots, Z_n$. As demonstrated in \cite{Roy:2012:robit}, the posterior density $\pi(\bb \mid {\bf z})$ is 
intractable, and \cite{Roy:2012:robit} develops a Data DA approach to construct a computationally tractable Markov chain with $\pi(\bb \mid {\bf y})$ as its stationary density by introducing unobserved latent variables $\{(U_i, \lambda_i)\}_{i=1}^n$ that are mutually independent and satisfy $U_i \mid \lambda_i \sim \mathcal{N}({\bf w}_i^T \bb, 1/{\bf \lambda}_i)$ and $\lambda_i \sim \mbox{Gamma}(\nu/2,\nu/2)$. Straightforward calculations now show that 
$U_i \sim t_\nu ({\bf w}_i^T\bb, 1)$, where $t_\nu (\mu, \sigma)$ denotes the Student's $t$-distribution with $\nu$ 
degrees of freedom, location $\mu$ and scale $\sigma$. Furthermore, if $Y_i = 1_{\{Z_i > 0\}}$, then $P(Y_i = 1 \mid \bb) = P(U_i > 0) = F_\nu 
({\bf w}_i^T \bb)$, which precisely corresponds to the robit regression model specified above. 

One can map this setting to the DA framework laid out in Section \ref{sec:intro}, by viewing $\bb$ as ``$X$", $({\bf U}, {\boldsymbol \lambda})$ as ``$Y$", and 
the joint posterior density of $\bb, {\bf U}, {\boldsymbol \lambda}$ as ``$f_{X,Y}$". Straightforward calculations (see \cite{Roy:2012:robit}) show that 
samples from $f_{X \mid Y}$ and $f_{Y \mid X}$ are easy to generate, as they involve sampling from standard distributions such as multivariate normal, 
truncated-$t$ and Gamma. These observations allow \cite{Roy:2012:robit} to use the corresponding DA Markov chain on $\mathbb{R}^p$ to generate samples from the 
$f_X$ (the marginal posterior density of $\bb$). We will refer to this Markov chain as the robit DA Markov chain. 

\cite{Roy:2012:robit} established Harris ergodicity of the robit DA chain, and investigated and established geometric 
ergodicity using a drift and minorization analysis. However, this analysis requires the following assumptions. 
\begin{itemize}
    \item The design matrix $W$ is full rank (which implies $n \geq p$ and rules out high-dimensional settings)
    \item $\Sigma_a = g^{-1} W^T W$ 
    \item $n \leq \frac{g^{-1} \nu}{(\nu+1)(1+2\sqrt{\bb_a^T W^T W \bb_a})}.$
\end{itemize}

\noindent
The last upper bound on $n$ involving the design matrix, the prior mean and covariance and the degrees of freedom $\nu$ is in particular very restrictive. Through a tighter drift and minorization analysis, \cite{mukh:khar:2023} relax the above restrictions to some extent, but not substantially (see Theorem S1 and Theorem S2 in \cite{mukh:khar:2023}). The related probit DA chain is obtained by (a) using the standard normal CDF $\Phi$ instead of $F_\nu$ in (\ref{robitmodel}), (b) using again a multivariate normal prior for $\bb$, (c) introducing latent variables $U_i \sim \mathcal{N}({\bf w}_i^T \bb, 1)$ with $Y_i = 1_{\{Z_i > 0\}}$ for 
$1 \leq i \leq n$, and (d) employing the DA machinery in Section \ref{sec:intro} with $(\bb, {\bf U})$ as ``$(X, Y)$". The geometric ergodicity analysis of the probit DA chain in \cite{roy:hobe:2007, chak:khar:2017} uses similar drift functions as in \cite{Roy:2012:robit, mukh:khar:2023} but requires minimal assumptions. Note that the latent variables $\{\lambda_i\}_{i=1}^n$ used in the robit setting are not needed in the probit setting. While having this additional layer of latent variables definitely complicates the analysis of the robit DA chain, it is not clear if the restrictive conditions listed above for geometric ergodicity of the robit DA chain are really necessary or if they are an artifact of using the drift and minorization technique in face of this added complication. 

A trace class analysis for the probit and robit DA chains is available in \cite{chak:khar:2017} and \cite{mukh:khar:2023} respectively. The trace class 
property for the probit DA chain was established in \cite[Theorem 2]{chak:khar:2017} under some constraints on $W$ and the prior covariance matrix 
$\Sigma_a$. This is consistent with the expectation that weaker assumptions should be needed for establishing drift and minorization based geometric 
ergodicity, as compared to establishing the trace class property. However, for the robit DA chain the reverse phenomenon holds: \cite{mukh:khar:2023} establish the 
trace class property for the robit DA chain {\it for any $n, \nu > 2, W, {\bf z}, \bb_a, \Sigma_a$}. Hence, drift and minorization approach, with the drift 
functions chosen in \cite{Roy:2012:robit, mukh:khar:2023}, needs stronger conditions to succeed than the trace class approach. Essentially, the additional layer of 
latent variables $\lambda_i$ introduced in the robit setting severely hampers the drift and minorization analysis, but does not cause additional complications for establishing the trace 
class condition. 

The key takeaway from the above discussion is that even if a successful drift and minorization based analysis is available, it is worth investigating the finiteness of the trace class integral in (\ref{integraltc}), it might require comparatively weaker assumptions contrary to expectations. Also, establishing the trace class property provides additional insights which are discussed in the next subsection. 

\subsection{Trace class property as a tool to compare DA and sandwich chains} \label{trace:comparison}

The results discussed previously show that, under certain regularity conditions, the sandwich chain is at least as good as the DA chain in certain respects. These results, while very useful, are not completely satisfactory as they only establish that the sandwich chain is ``as good as" the DA chain. The question is - are there any conditions under which the sandwich chain can be shown to be ``strictly better" 
than the DA chain (in an appropriate sense)? This question has been investigated in \cite{khar:hobe:2011, roy:2012a}, and we will discuss their results below. 

First, we provide a brief review of some ideas and results used in these analyses. 
Both papers leverage the fact (see \cite{Buja:1990, diac:khar:salo:2008}) that the DA 
operator $K$ can be written as a product of two simple `projection' operators. In particular, let $P_X$ denotes the operator from $L^2_0 (f_Y)$ to $L^2_0 (f_X)$ defined by 
$$
(P_X h) (x) = \int_{\mathbb{R}^q} h(y) f_{Y \mid X} (y \mid x) dy
$$

\noindent
for every $h \in L^2_0 (f_Y)$, and $P_Y$ denotes the operator from $L^2_0 (f_X)$ to $L^2_0 (f_Y)$ defined by 
$$
(P_Y g) (y) = \int_{\mathbb{R}^p} g(x) f_{X \mid Y} (x \mid y) dx
$$

\noindent
for every $g \in L^2_0 (f_X)$. Also, let $R: L^2_0 (f_Y) \rightarrow L^2_0 (f_Y)$ denotes 
the operator corresponding to the Mtd $r$ (the sandwich step Mtd). Then, it can be shown 
that 
$$
K = P_X P_Y \quad \mbox{ and } \quad K_{SA} = P_X R P_Y. 
$$

\noindent
For any $g \in L^2_0 (f_X)$ and any $h \in L^2_0 (f_Y)$, simple calculations show that 
$<P_X h, g> = <h, P_Y g>$ (the first inner product corresponds to $L^2_0 (f_X)$, while the 
second one corresponds to $L^2_0 (f_Y)$). Hence, {\it the operators $P_X$ and $P_Y$ are 
adjoints of each other}. Using elementary operator theoretic arguments, this fact can be used to immediately 
conclude that $K = P_X P_Y$ is a positive operator (as previously mentioned) and $\|K\| = 
\|P_X\|^2 = \|P_Y\|^2$. It follows that 
$$
\|K_{SA}\| = \|P_X R P_Y\| \leq \|P_X\| \|R\| \|P_Y\| = \|R\| \|K\|. 
$$

\noindent
However, in almost all applications of interest, $\|R\| = 1$, since the sandwich step corresponds typically to a univariate movement in $\mathbb{R}^q$. Hence, the above argument does not allow us to prove that the sandwich algorithm is strictly better (in the norm sense). 

On the other hand, if the trace class condition in (\ref{integraltc}) is satisfied, then a very useful singular value decomposition of $f_{X,Y}$ becomes available (see \cite{Buja:1990}). In particular, it can be shown that 
$$
\frac{f_{X,Y} (x,y)}{f_X (x) f_Y (y)} = \sum_{i=0}^\infty \beta_i g_i (x) h_i (y), 
$$

\noindent
where $\{g_i\}_{i=1}^\infty$ and $\{h_j\}_{j=1}^\infty$ form orthonormal bases of $L^2_0 (f_X)$ and $L^2_0 (f_Y)$ respectively (with $g_0 \equiv 1, h_0 \equiv 1$), $\{\beta_i\}_{i=0}^\infty$ is a decreasing sequence of numbers in the unit interval with $\beta_0 = 1$, and $g_i$ and $h_j$ are orthogonal for every $i \neq j$ in the sense that 
$$
\int_{\mathbb{R}^p} \int_{\mathbb{R}^q} g_i (x) h_j (y) f_{X,Y} (x,y) dy dx = 0. 
$$

\noindent
The singular value decomposition can be used to establish that 
\begin{equation} \label{singular}
P_Y g_i = \beta_i h_i \quad \mbox{ and } P_X h_i = \beta_i g_i 
\end{equation}

\noindent
for every $i \geq 0$. In particular this implies $\{\beta_i^2\}_{i=1}^\infty$ is the spectrum of the DA operator $K$ and $\{g_i\}_{i=1}^\infty$ 
are the corresponding eigenfunctions. Given that $K_{SA} = P_X R P_Y$ and (\ref{singular}), a comparative spectral analysis for the 
DA and sandwich algorithms now hinges on how the operator $R$ interacts with the functions $\{h_i\}_{i=1}^\infty$. By investigating this carefully, \cite{khar:hobe:2011} establish the following result. 
\begin{theorem} \label{trace:class:comp}
Assume that (\ref{integraltc}) holds, and that the sandwich operator $R$ is idempotent ($R^2 = R$) with $\|R\| = 1$. Then the sandwich operator $K_{SA}$ is positive and trace class. If $\{\lambda_{SA,i}\}_{i=1}^\infty$ denotes the ordered sequence of eigenvalues for $K_{SA}$, then $\lambda_{SA,i} \leq \beta_i^2$ for every $i \geq 1$. Also, for every $i$ such that $\beta_i > 0$, $\lambda_{SA,i} = \beta_i^2$ {\it if and only if} $Rh_i = h_i$. 
\end{theorem}

\noindent
In other words, the spectrum of the sandwich operator $K_{SA}$ is dominated {\it pointwise} by the spectrum of the DA operator $K$, and strict domination for a specific pair of eigenvalues depends {\it exclusively} on whether or not the operator $R$ leaves the corresponding $h$-function invariant. A stronger version of this result, which only requires 
$K$ to be compact is established in \cite{roy:2012a}. 

For a more specific comparison of the operator norms of the two operators, as expected the key factor is the interaction of the the operator $R$ with the linear span of $\{h_i\}_{i=1}^\ell$, where $\ell$ is the multiplicity of the largest eigenvalue $\beta_1$, i.e., number of $\beta_i$'s that are exactly equal to $\beta_1$. The next result from \cite{khar:hobe:2011} leverages this idea to provide a necessary and sufficient condition for the operator norm of $K_{SA}$ to be strictly smaller than the operator norm of $K$. 
\begin{theorem}
    Under the setup in Theorem \ref{trace:class:comp}, $\|K_{SA}\| < \|K\|$ if and only if 
    the only function in the linear span of $\{h_i\}_{i=1}^\ell$ which is left invariant by $R$ is the (identically) zero function. 
\end{theorem}

\noindent
While the two results above provide necessary and sufficient conditions for the equality of the norm (and other eigenvalues) for $K$ and $K_{SA}$, their practical utility is somewhat limited by the requirement to identify the functions $\{h_i\}_{i=1}^\infty$. However, if we restrict to the group action based Haar PX-DA chain, then the following result can be established. 
\begin{theorem} \label{haar:trace}
Assume that (\ref{integraltc}) holds, and that the operator $K_{SA}$ corresponds to the Haar PX-DA sandwich chain outlined in Section \ref{sec:imprda}. Then $\lambda_{i,SA} = \lambda_i$ for all $i \geq 1$ if and only if 
$$
f_{X \mid Y} (x \mid y) = f_{X \mid Y} (x \mid gy) \; \forall g \in G, x \in \mathbb{R}^p, y \in \mathbb{R}^q. 
$$
\end{theorem}

\noindent
If the condition in (\ref{haar:trace}) is satisfied, then it can be shown that the Mtds $k$ and $k_{SA}$ {\it are exactly the same}. Hence, outside of this triviality, the Haar PX-DA sandwich chain is strictly better than the DA chain in the sense that its eigenvalues are uniformly (pointwise) dominated by those of the DA chain with at least one strict domination. 


\section{The ``two-block" DA algorithm and its sandwich variants} \label{sec:twoblock}

The applicability of the DA algorithm depends on the ease of sampling from the conditional densities $f_{Y \mid X}$ and $f_{X \mid Y}$. While samples from both of these densities can be generated in a straightforward way in various settings, there are many others where this is not true. In particular, in these settings the density $f_{X \mid Y}$ is often intractable and cannot be directly sampled from. However, it is often possible to partition 
$X$ into two components $X = (U, V)$ (with $U \in \mathbb{R}^u$ and $V \in \mathbb{R}^{p-u}$) such that samples from $f_{U \mid V, Y}$ and $f_{V \mid U, Y}$ can be easily generated. In such settings, a modified {\it two-block DA algorithm} is used to generate samples from $f_X = f_{U,V}$. Each iteration of the DA algorithm consists of three steps --- a draw
from $f_{Y|X}$ followed by a draw from $f_{U|V,Y}$, and finally a draw from $f_{V|U, Y}$. The transition of this two-block DA Markov chain from the current state $x = (u,v)$ to the next state $x' = (u', v')$ can be described as follows.
\begin{algorithm}[H]
  \caption{One-step transition of the two-block DA Markov chain}
  \label{alg:da:tb}
  \begin{algorithmic}[1]

    \vspace{.1in}
    
    \STATE Given $x = (u,v)$, generate a draw from $f_{Y|X}(\cdot|x)$, and call the observed value $y$.

    \vspace{.1in}
    
    \STATE Generate a draw from $f_{U|V,Y}(\cdot|v,y)$, and call the observed value $u'$. 

    \vspace{.1in}
    
    \STATE Generate a draw from $f_{V|U,Y}(\cdot|u',y)$, and call the observed value $v'$. 
    
    \vspace{.1in}
    
\end{algorithmic}
\end{algorithm}

\noindent
From the three steps in each iteration of the two-block DA algorithm, it follows that the Markov 
transition density (Mtd) of the DA Markov chain $\{ \tilde{X}_n \}_{n\ge 1}$ (with 
$\tilde{X}_n = (U_n, V_n)$) is given by 
\begin{equation}
  \label{eq:da_mtd1}
  k_{TB} (x'|x) = \int_{\mathbb{R}^q} f_{V|U,Y}(v'|u',y) f_{U|V,Y}(u'|v,y) f_{Y|X}(y|x) \, dy \;.
\end{equation}

\noindent
Its easy to show that the two-block DA chain has $f_X$ as its stationary density. However, a 
key difference as compared to the ``single-block" Markov chain $\{X_n\}_{n \geq 0}$ in Section~\ref{sec:intro} 
is that the two-block DA chain {\it does not} satisfy the detailed balance 
condition. 

\medskip

\noindent
{\bf Example (Bayesian quantile regression)} Consider the linear model 
$$
Z_i = {\bf w}_i^T \bb + \epsilon_i ; i=1,2,...,n 
$$

\noindent
subject to the constraint that the $\alpha^{th}$ quantile of the error distribution of is $0$. Recall that $\{{\bf w}_i\}_{i=1}^n$ is the collection of $p$-
dimensional covariate vectors, and $\bb \in \mathbb{R}^p$ are the regression coefficients. The standard (frequentist) method for estimating the 
regression parameter $\bb$ (see \cite{yu:moye:2001}) minimizes an objective function which is equivalent to the negative log-likelihood for $\bb$ if the 
errors are assumed to be i.i.d. with the asymmetric Laplace density given by 
$$
g_\alpha (\epsilon) = \alpha(1-\alpha) e^{-\rho_\alpha (\epsilon)}. 
$$

\noindent
where $\rho_\alpha (x)=x\left( \alpha-I(x<0) \right)$ with $I(\cdot)$ denoting the standard indicator function. Note that $g_\alpha$ has $\alpha^{th}$ quantile 
equal to zero, and corresponds to the standard Laplace density with location and scale equal to $0$ and $1/2$, respectively, when $\alpha = 1/2$. 
Leveraging this analogy, \cite{yu:moye:2001, kozu:koba:2011} pursue a Bayesian approach where $\{\epsilon_i\}_{i=1}^n$ are assumed to be i.i.d. with common density $g_\alpha$, with $\sigma > 0$ is an unknown scale parameter. The following independent priors are assigned to $\bb$ and $\sigma$: $\bb \sim \mathcal{N}_p (\beta_0, B_0)$ and $\sigma \sim IG \left( \frac{n_0}{2}, \frac{t_0}{2} \right)$. The posterior density of $(\bb, \sigma)$ is intractable, and 
\cite{kozu:koba:2011} propose the following DA approach which exploits a normal/exponential mixture representation of the asymmetric Laplace distribution.

\noindent
Define $\theta := \theta(\alpha) = \frac{1-2\alpha}{\alpha(1-\alpha)}$ and $\tau^2 := \tau^2 (\alpha) = 
\frac{2}{\alpha(1-\alpha)}$, and consider random pairs $\left\{(Z_i, R_i)\right\}_{i=1}^{n}$ such that 
$Z_i \mid R_i = r_i, \bb, \sigma \sim \mathcal{N} ({\bf w}_i^T \bb + \theta r_i, r_i \sigma \tau^2)$ and $R_i \mid \bb, \sigma \sim Exp(\sigma)$. It can be 
easily verified that the marginal density of $\frac{Z_i - {\bf w}_i^T \bb}{\sigma}$ given $(\bb, \sigma)$ is indeed $g_\alpha$. However, direct sampling or 
closed form computations for the joint posterior density of $\bb, {\bf R}, \sigma$ given ${\bf Z} = {\bf z}$ are not feasible. However, it can be shown that 
(a) the full posterior conditional distribution of $\bb$ is multivariate Gaussian, (b) the elements of ${\bf R}$ are conditionally independent given $\bb, 
\sigma, {\bf Z} = {\bf z}$, and follow a generalized inverse Gaussian distribution, (c)  the full posterior conditional distribution of $\sigma$ is inverse gamma. 

A two-block DA algorithm with $\bb$ as ``$U$", ${\bf R}$ as ``$V$" and $\sigma$ as ``$Y$" can hence be used to generate approximate samples from the joint 
posterior density of $(\bb, {\bf Z})$. This in particular yields samples from the marginal posterior density of $\bb$. Since the posterior density of $\sigma$ given only $\bb$ can be shown to be an inverse gamma density, this provides a mechanism to sample from the target posterior density of $(\beta, \sigma)$. 


Coming back to the general setting, a direct application/insertion of the single-block Haar PX-DA step (Step 2) as a sandwich step in the two-block setting is not feasible, as 
the resulting Markov chain does not in general have $f_X$ as its stationary density. 
\cite{pal:khare:2015} develop feasible two-block adaptations of the single-block Haar PX-DA sandwich algorithm in Section \ref{sec:imprda}, which we briefly describe below. 

Consider a group $G$ acting topologically on the left of $\mathbb{R}^q$, as in the 
discussion prior to the single-block Haar PX-DA algorithm. Given any $x \in \mathbb{R}^p, y \in 
\mathbb{R}^q$, define the densities 
$$
f_{x,y}^1 (g) = \frac{f_{V,Y}(v,gy) \chi(g)}{C_1 (x,y)} \; \forall g \in G
$$

\noindent
and 
$$
f_{x,y}^2 (g) = \frac{f_{U,V,Y}(u,v,gy) \chi(g)}{C_2 (x,y)} \; \forall g \in G. 
$$

\noindent
Here $C_1 (x,y)$ and $C_2 (x,y)$ (assumed to be finite) are normalizing constants. Either of 
these two densities can be used to construct a 'valid' Haar PX-DA sandwich step in the 
current setting. The transition of this two-block sandwich Markov chain from the current 
state $x = (u,v)$ to the next state $x' = (u', v')$ can be described as follows. 
\begin{algorithm}[H]
  \caption{One-step transition of the two-block Haar PX-DA sandwich Markov chain}
  \label{alg:px:tb}
  \begin{algorithmic}[1]

    \vspace{.1in}
    
    \STATE Given $x = (u,v)$, generate a draw from $f_{Y|X}(\cdot|x)$, and call the observed value $y$. 

    \vspace{.1in}
    
    \STATE Generate a draw from $f_{x,y}^j$ (for $j=1$ or $j=2$) and call the observed value $g$. 

    \vspace{.1in}
    
    \STATE Generate a draw from $f_{U|V,Y}(\cdot|v,gy)$, and call the observed value $u'$. 

    \vspace{.1in}
    
    \STATE Generate a draw from $f_{V|U,Y}(\cdot|u',gy)$, and call the observed value $v'$. 
    
    \vspace{.1in}
    
\end{algorithmic}
\end{algorithm}

Let $k_{TB,SA1}$ and $k_{TB,SA2}$ denote the Mtds corresponding to the two-
block sandwich chain described in two-block Haar PX-DA algorithm (with $j=1$ and $j=2$ 
respectively). \cite{pal:khare:2015} show that both $k_{TB,SA1}$ and $k_{TB,SA2}$ 
have $f_X$ as their stationary distribution. 

Returning to the Bayesian quantile regression example, recall that the the scale parameter $\sigma$ plays the role of "$Y$". Consider the action of the group $G = \mathbb{R}_+$ on $\mathbb{R}_+$, the sample space of $\sigma$, through scalar multiplication. The left Haar  measure for $G$ is given by $\nu_l (dg) = \frac{dg}{g}$, and it can be shown that the Lebesgue measure on $\mathbb{R}_+$ is relatively invariant with respect to the multiplier $\chi(g) = g$. Using the above construction of two-block sandwich densities, it can be shown that $f^1_{x,y}$ is a non-standard univariate density. However, samples can be generated from this density using a rejection sampler with a dominating inverse gamma density. On the other hand, $f^2_{x,y}$ can be shown to be an inverse gamma density, and the corresponding sandwich chain is more suitable for practical use. 


We conclude with a quick summary of associated theoretical results for the two-block DA setting. Using results in \cite{asmu:glyn:2011}, it follows that the two-block 
DA chain and the sandwich chain are Harris ergodic if the corresponding Mtd 
is strictly positive everywhere (this is the case with most chains 
encountered in statistical applications). However, things get much more complicated (compared to the single-block setting) for a comparative study of deeper issues such as operator norms, geometric ergodicity etc. A key reason for these challenges is that the two-block DA operator, based on the three-stage transition step, is a product of three relevant projection operators, and consequently loses its self-adjointness and positivity. A theoretical comparison of closely related ``lifted" versions of the two-block DA and sandwich chains is available in \cite{pal:khare:2015}, however several questions involving a direct comparison of the two-block DA and sandwich chains remain open. 
\begin{itemize}
\item Under what conditions does geometric ergodicity of the two block DA chain imply geometric ergodicity of the corresponding sandwich chain? 
\item Although reversibility is lost, one could still aim to establish that the two-block DA operator is Hilbert-Schmidt by showing that the condition in (\ref{integralhs}) is satisfied. In such a setting, under what conditions does the Hilbert-Schmidt property of the 
two-block DA chain imply the same for the corresponding sandwich chain? 
\item How do answers to the above questions depend on the choice of $j=1$ vs. $j=2$ in the sandwich step in the two-block Haar PX-DA algorithm? 
\end{itemize}


\section{Distributed DA adaptations for massive data settings}
\label{sec:adda}

\noindent
The sandwich DA algorithm can serve as a useful and computationally inexpensive tool to speed up the convergence of the DA algorithm. However, 
its utility is limited in increasingly common ``massive" datasets, which contain hundreds of thousands (if not millions) of samples and/or variables. For a Bayesian statistical analysis of such datasets, a key goal is to generate samples from the intractable posterior density $f_X$ of the parameters ``$X$".  For instances where the DA algorithm is applicable, the dimensionality of ``$Y$" (the latent variables) is often equal to the number of samples or the number of variables. For example, in the Bayesian variable selection example described in Section \ref{sec:bvs}, the dimensionality of $Y$ is $p$, the number of variables. The corresponding methodology is meant to be used for high-dimensional settings where $p$ is very large. In such settings, the computational cost of sampling the latent variables in each iteration of the DA and sandwich chains can be exorbitant. Alternatively, consider the Bayesian logistic regression example in Section \ref{sec:logit}, where the number of latent variables is equal to the number of samples $m$. For datasets such as the MovieLens Data \cite{Perry:2016, SDL:2019}, where the number of samples is in the millions, the computational cost of sampling the latent variables can again be prohibitive. 

For settings where a massive number of samples is an issue, the following divide-and-conquer strategy can be considered: (a) Divide the data into a number of smaller subsets of reasonable size (b) run DA Markov chains on all subsets in parallel, and (c) suitably combine the different subset-wise Markov chains for inference. See \cite{doi:10.1080/17509653.2016.1142191, JLY:2019, wang2021divideandconquer} and the references therein. The combined draws from the subset-wise Markov chains do not constitute a Markov chain, and quantification of the Monte Carlo error becomes challenging. While asymptotic statistical guarantees based on a normal approximation of the target posterior are available for some of these methods (as the subset sample size tends to infinity), no rigorous bounds for the distance between the distribution of the combined parameter draws and the target posterior distribution (based on the entire data) are available. In other words, the lack of a Markov chain structure for the combined draws deprives the user of the rich diagnostic and theoretical tools that are available in the literature for Markov chains. 

To address the above issues, the authors in \cite{ZKS:2023} develop an {\it asynchronous} and {\it distributed} version of the DA algorithm, called {\it ADDA}, which outputs a Markov chain with the desired posterior as its stationary distribution. Two key assumptions that are needed for ADDA to be useful are - (a) the latent variables can be partitioned into blocks which are conditionally independent (given the original parameters ``$X$" and the data), and (b) the conditional posterior distribution of each latent sub-block exclusively depends on a relevant subset of the observed data. This is for example true for the Bayesian variable selection example in Section \ref{sec:bvs}, where the augmented variables $\{\tilde{Y}_j\}_{j=1}^p$ are conditionally independent given the regression coefficients and the data. Also, the conditional posterior distribution of $\tilde{Y}_j$ does not depend on the observed data (hence, requirement (b) above is trivially satisfied). Again, for the Bayesian logistic regression example in Section \ref{sec:logit}, the augmented variables $\{\tilde{Y}_i\}_{i=1}^m$ are conditionally independent given the regression coefficients and the data, and the conditional posterior distribution of $\tilde{Y}_i$ depends only on $(\ell_i, w_i)$ (quantities related to the $i^{th}$ observation). 

Consider the general DA setup of Section \ref{sec:intro}, where $X$ represents the parameter of interest, and $Y$ represents the block of latent parameters. The goal is to sample from $f_X$, which represents the marginal posterior density of $X$ given the observed data (the conditioning on the observed data will be left implicit for continuity and ease of exposition). We assume that $Y$ can be divided into $k$ sub-blocks $Y^1, Y^2, \cdots, Y^k$ which are conditionally independent given $X$. Also, as mentioned above, the observed data can be partitioned into $k$ subsets such that $f_{Y^j \mid X}$ only depends on the $j^{th}$ data subset for $j=1,2, \cdots, k$. 

The ADDA algorithm employs $1$ manager process and $k$ worker processes. At the level of sampling, the job of the manager process is to sample new parameter values from $f_{X \mid Y}$ when appropriate, and to send this sample to all the workers. The job of the $j^{th}$ worker process is to sample new values for the $j^{th}$ latent block from $f_{Y^j \mid X}$ when appropriate, and to send these samples to the manager process. Hence, the job of sampling the latent parameter $Y$ is {\it distributed} to the $k$ worker processes. Another key feature of the ADDA algorithm is {\it asynchronous} sampling of the $k$ latent parameter blocks. The degree of asynchrony is controlled by user-specified parameters $r \in (0,1]$ and $\epsilon \in (0,1]$ as follows. To make its next conditional draw of the parameter $X$, the manager process, with probability $\epsilon$, waits to receive updated draws of the relevant latent parameter blocks from all the workers. But with probability $1-\epsilon$, it only waits to receive an $r$-fraction of updated draws, and proceeds to sample $X$ given the most recent draws for all latent parameter blocks. If a worker is in the midst of sampling its assigned latent parameter block, but receives a new $X$-draw from the manager, it 
terminates its current sampling, and begins afresh using the latest $X$-draw received from the manager. This entire process is summarized below. 
\begin{enumerate}[label={(\arabic*)},ref=(\arabic*)]
\item At time $t=0$, the manager starts with initial values $(\tilde{X}_0, \tilde{Y}_0)$ at $t=0$ and sends $X_0$ to the workers. 
\item For $t=0, 1, \ldots, \infty$, the manager 
  \begin{enumerate}[label={(M-\alph*)},ref=(\alph*), noitemsep, topsep=0pt ]
  \item waits to receive only an $r$-fraction of updated $\tilde{Y}_{t+1}^j$s (see below) from the workers with probability $1-\epsilon$, and with probability $\epsilon$, waits to receive all the updated $\tilde{Y}_{(t+1)}^j$s from the workers;
  \item creates $\tilde{Y}_{t+1}$ by replacing the relevant $\tilde{Y}^j_t$s with the newly received $\tilde{Y}^j_{t+1}$;
  \item draws $X_{t+1}$ from $f(X \mid \tilde{Y}_{t+1})$; and 
  \item sends $X_{t+1}$ to all the worker processes and resets $t = t+1$.  
  \end{enumerate}
\item For $t=0, \ldots,\infty$, the worker $j$ ($j=1, \ldots, k$) 
  \begin{enumerate}[label={(W-\alph*)},ref=(\alph*), noitemsep, topsep=0pt ]
  \item waits to receive $X_t$ from the manager process;
  \item draws $\tilde{Y}^j_{t+1}$ from $p(Y^j \mid \tilde{X}_{t+1})$;   and
  \item sends $\tilde{Y}^j_{t+1}$ to the manager process, resets $t = t+1$, and goes to (W-a) if $\tilde{X}_{t+1}$ is not received from the manager before the draw is complete; otherwise, it truncates the sampling process, resets $t = t+1$, and goes to (W-b).
  \end{enumerate}
\end{enumerate}

\noindent
It is clear that when $\epsilon = 1$ or $r = 1$, the ADDA algorithm is identical to the DA algorithm. However, things get interesting when $\epsilon < 1$ and 
$r < 1$. In this setting, at each iteration of the ADDA algorithm, only an $r$-fraction of the latent parameter blocks are updated (with probability $1-
\epsilon$). The ADDA algorithm can then be viewed as a mix of systematic and random subset scan Gibbs sampler. The systematic part comes from always updating 
$X$ in any given iteration, and the random subset part comes from only updating a random subset of $[kr]$ sub-blocks of $Y$ at each iteration. Such algorithms 
have been previously considered in the literature. However, what gives the ADDA a novel and interesting flavor is that unlike existing approaches, the 
comparative speed of the $k$ workers for sampling the respective $Y$ sub-blocks can depend on the current value of $X$. In other words, the choice of the 
$[kr]$ sub-blocks of $Y$ that are updated at any given iteration {\it can depend on the current value of the parameter $X$}. Given this dependence, it turns 
out that the marginal $X$-process $\{\tilde{X}_t\}_{t \geq 0}$ is not Markov, but the joint $(X,Y)$-process $\{\tilde{X}_t, \tilde{Y}_t\}_{t \geq 0}$ is 
Markov. 

In fact, it can be shown that the Markov chain $\{\tilde{X}_t, \tilde{Y}_t\}_{t \geq 0}$ has 
$f_{X,Y}$ as its stationary (invariant) density. For ease of exposition, we provide a proof of this 
assertion when $Y$ and $X$ are supported on a discrete space and $k$, the number of latent variable 
blocks, is equal to $2$. We assume that $r = 0.5$ and $\epsilon = 0$. The arguments below can be 
extended in a straightforward way to a general setting with more than two latent variable blocks, 
and to non-discrete settings with arbitrary $r$ and $\epsilon$ values in $[0,1]$. 

Let $(Y_0, X_0)$ denote the starting value for the ADDA chain, and assume that it is drawn from the 
desired posterior $f_{Y, X}$. Let $(Y_1, X_1)$ denote the next iterate generated by the ADDA chain. 
Our goal is to show that $(Y_1, X_1) \sim f_{Y, X}$. As we will see below, {\it a key assumption 
which enables this is the conditional independence assumption which implies $f_{Y \mid X} (\tilde{y} 
\mid \tilde{x}) = f_{Y^1 \mid X} (\tilde{y}_1 \mid \tilde{x}) f_{Y^2 \mid X} (\tilde{y}_2 \mid 
\tilde{x})$} (recall that $\tilde{y}_1$ and $\tilde{y}_2$ denote the two blocks of $\tilde{y}$). 

Since $X_1$ given $Y_1$ is a draw from $f_{X \mid Y}$, it follows that 
\begin{eqnarray*}
P((Y_1, X_1) = (\tilde{y}, \tilde{x})) 
&=& P(X_1  = \tilde{x} \mid Y_1 = \tilde{y}) P(Y_1 = \tilde{y})\\
&=& f_{X \mid Y} (\tilde{x} \mid \tilde{y}) P(Y_1 = \tilde{y}). 
\end{eqnarray*}

\noindent
Hence, to prove the desired result, it is enough to show that $P(Y_1 = \tilde{y}) = f_{Y} (\tilde{y})$. Note that 
$$
P(Y_1 = \tilde{y}) = \sum_{y', x'} P(Y_1 = \tilde{y} \mid (Y_0, X_0) = (y', x')) 
f_{Y, X} (y', x'). 
$$

\noindent
Let us recall how $Y_1$ is sampled given $(Y_0, X_0) = (y', x')$. With probability say $c_1 
(x')$, only the first latent variable block $Y^1_1$ is obtained using a draw from $f_{Y^1 \mid X} (\cdot 
\mid x')$ and the second block is left unchanged at $y_2'$, and with probability $c_2 (x') = 1 - c_1 (x')$, only 
the second latent variable block $Y^2_1$ is obtained using $f_{Y^2 \mid X} (\cdot \mid x')$ and the 
first block is left unchanged at $y_1'$. It follows that 
\begin{eqnarray*}
& & P(Y_1 = \tilde{y})\\
&=& \sum_{y', x'} c_1 (x') f_{Y^1 \mid X} (\tilde{y}_1 \mid x') 1_{\{y_2' = \tilde{y}_2\}} f_{Y, 
X} (y', x') + \sum_{y', x'} c_2 (x') f_{Y^2 \mid X} (\tilde{y}_2 \mid x') 1_{\{y_1' = \tilde{y}_1\}} 
f_{Y, X} (y', x')
\end{eqnarray*}

\noindent
Using $f_{Y, X} (y', x') = f_{Y^1 \mid X} (y_1' \mid x') f_{Y^2 \mid X} (y_2' 
\mid x') f_X (x')$ (by conditional independence of $Y^1$ and $Y^2$ given $X$), we get 
\begin{eqnarray*}
& & P(Y_1 = \tilde{y})\\
&=& \sum_{x'} \sum_{y_1'} \sum_{y_2'} c_1 (x') f_{Y^1 \mid X} (\tilde{y}_1 \mid x') 1_{\{y_2' = \tilde{y}_2\}} 
f_{Y^1 \mid X} (y_1' \mid x') f_{Y^2 \mid X} (y_2' \mid x') f_X (x') +\\
& & \sum_{x'} \sum_{y_1'} \sum_{y_2'} c_2 (x') f_{Y^2 \mid X} (\tilde{y}_2 \mid x') 1_{\{y_1' = \tilde{y}_1\}} 
f_{Y^1 \mid X} (y_1' \mid x') f_{Y^2 \mid X} (y_2' \mid x') f_X (x')\\
&=& \sum_{x'} c_1 (x') f_{Y^1 \mid X} (\tilde{y}_1 \mid x') f_X (x') \left( \sum_{y_1'} 
f_{Y^1 \mid X} (y_1' \mid x') \right) \left( \sum_{y_2'} f_{Y^2 \mid X} (y_2' \mid x') 1_{\{y_2' = 
\tilde{y}_2\}} \right) +\\
& & \sum_{x'} c_2 (x') f_{Y^2 \mid X} (\tilde{y}_2 \mid x') f_X (x') \left( \sum_{y_2'} 
f_{Y^2 \mid X} (y_2' \mid x') \right) \left( \sum_{y_1'} f_{Y^1 \mid X} (y_1' \mid x') 1_{\{y_1' = 
\tilde{y}_1\}} \right)\\
&=& \sum_{x'} c_1 (x') f_{Y^1 \mid X} (\tilde{y}_1 \mid x') f_X (x') f_{Y^2 \mid X} 
(\tilde{y}_2 \mid x') + \sum_{x'} c_2 (x') f_{Y^2 \mid X} (\tilde{y}_2 \mid x') f_X (x') 
f_{Y^1 \mid X} (\tilde{y}_1 \mid x')\\
&=& \sum_{x'} (c_1 (x') + c_2 (x')) f_{Y, X} (\tilde{y}, x'). 
\end{eqnarray*}

\noindent
The last step again uses conditional independence of $Y^1$ and $Y^2$ given $X$. Since $c_1 (x') + 
c_2 (x') = 1$, it follows that 
$$
P(Y_1 = \tilde{y}) = \sum_{x'} f_{Y, X} (\tilde{y}, x') = f_{Y} (\tilde{y}). 
$$

\noindent
Note that the above argument considers the pure asynchronous ADDA ($\epsilon = 0$). But the ADDA kernel with positive $\epsilon$ is 
a mixture of the DA and pure asynchronous ADDA kernels, so the result immediately follows for such settings as well. For a setting with 
more than two blocks and a general value of $r \in (0,1)$, we will have $J = {{K}\choose{\lceil Kr \rceil}}$ terms in the derivation above 
instead of two terms. The $j^{th}$ term, with essentially the same manipulations as above, will simplify to $c_j (x') f_{Y, 
X} (\tilde{y}, x')$, where $c_j (x')$ denotes the probability of choosing the relevant subset of latent variable blocks. Since 
$\sum_{j=1}^J c_j (x') = 1$, the invariance result will follow. The authors in \cite{ZKS:2023} also establish the geometric ergodicity of the ADDA chain in specific 
settings, including the Bayesian variable selection example considered in Section \ref{sec:bvs}. 

The parameter $r$ controls degree of asynchrony. If $r$ is small, then the computational cost of each ADDA iteration is lower, but the ADDA Markov chain mixes 
at a slower pace. This tradeoff between slow mixing and lower computational cost per iteration is key to the choice of $r$ and the performance of the ADDA 
algorithm. The hope is that for a range of values of $r$ where the joint effect of these two competing factors leads to a significant decrease in the overall 
wall clock time required for the ADDA chain (as compared to a pure distributed implementation of the original DA chain). This is indeed seen in the extensive experimental evaluations performed in \cite{ZKS:2023}, where the ADDA is shown to have a remarkable overall computational gain with a comparatively small loss of accuracy. For example, when analyzing the MovieLens data \cite{Perry:2016, SDL:2019} (with millions of samples) using Bayesian logistic regression (Section \ref{sec:logit}), the ADDA algorithm is three to five times faster with only 2\% less accuracy compared to the (pure distributed) DA algorithm after 10,000 iterations. 



\section{Summary and discussion}
\label{sec:summ}

DA algorithms introduce appropriate latent variables to facilitate
sampling from intractable target distributions. These algorithms
explore the parameter space by iteratively updating the
parameters and the latent variables, usually drawing from some
standard distributions. Indeed, DA algorithms are a popular choice for
MCMC computing due to its simplicity and broad applicability. The
simple structure of the transition density of the DA Markov chain
allows tractable detailed spectral analyses of its convergence
behavior in addition to drift and minorization based analyses. This chapter discusses the tools for analyzing and comparing the
spectrum of DA Markov chains. The applicability of DA methods has been
demonstrated by their successful implementation in various widely used
examples from statistics and machine learning.

Despite its widespread use, the DA algorithms can suffer from slow
mixing. A great deal of effort has gone into developing techniques to
tweak the DA chain's transition density to improve its convergence speed. Among the different approaches, parameter expansion based
methods, the PX-DA strategies, have been the most successful.
As described in this chapter, several theoretical results have been established
showing the superiority of the PX-DA algorithms over the DA methods. For a given
reparameterization, the results in the literature identify the `best' PX-DA algorithm
as the Haar PX-DA based on the Haar measure. 
On the other hand, there is not much study available on comparison among
different parameter expansion strategies. One exception is
\cite{roy:2014}, which
through an empirical study involving a Bayesian robit model, shows that a partially reparameterized Haar PX-DA
algorithm can outperform a fully reparameterized Haar PX-DA algorithm.
Using the same robit example, \cite{roy:2014} shows that some reparameterization
may provide little improvement over the DA algorithm. An important area of
future research is to provide theoretical and methodological results
for constructing and comparing different reparameterization strategies for improving
DA algorithms.


In various fields and applications, it is common to
observe large data sets that need to be analyzed. For these data sets, the number of observations or variables is large. The dimension of the latent vector $y$ in a DA scheme is
usually the same as the number of observations or the number of
variables. On the other hand, generally, as seen in all examples
considered in this chapter, the latent variables are conditionally
independent given the observed data and the current parameter
value. Thus, the draw from $y$, the first of the two steps in every
iteration of the DA algorithm, can be parallelized, distributing the
computational burden across multiple processors or nodes. Indeed,
conditional independence of the latent blocks $Y^1, Y^2, \cdots, Y^k$
is key for ensuring that the ADDA-process
$\{(\tilde{X}_t, \tilde{Y}_t)\}_{t=0}^\infty$ described in
Section~\ref{sec:adda} is a Markov chain with $f_{X,Y}$ as its
stationary distribution. However, such conditional independence is lost in many applications, such as hidden Markov models and time series models. An important open research direction is the extension of ADDA-type ideas that adapt to
this loss of conditional independence without compromising on some
underlying Markov structure. Another open direction is a useful integration of the
sandwich technique to develop parameter-expanded versions of the ADDA algorithm.

\section*{Acknowledgments}
 The first author’s work was partially supported by USDA NIFA Grant 2023-70412-41087.

\let\cleardoublepage\clearpage

\bibliographystyle{apalike} 
\bibliography{references}

\end{document}